\documentclass[twocolumn,showpacs,preprintnumbers,amsmath,amssymb]{revtex4}
\usepackage{graphicx}
\usepackage{dcolumn}
\usepackage{bm}

\begin{document}


\title{The High-Flux Backscattering Spectrometer at the NIST Center
  for Neutron Research}

\author{A.~Meyer$^{1,2}$}\altaffiliation[Now at ]{Physik Department E\,13, Technische Universit\"{a}t
M\"{u}nchen, 85747 Garching, Germany}

\author{R.\,M.~Dimeo$^1$}
\author{P.\,M.~Gehring$^1$}
\author{D.\,A.~Neumann$^1$}\email[Corresponding author: ]{dan@nist.gov}

\affiliation{$^1$National Institute of Standards and Technology, NIST
  Center for Neutron Research, Gaithersburg, Maryland 20899-8562 \\
  $^2$University of Maryland, Department for Materials and Nuclear
  Engineering, College Park, Maryland 20742}

\homepage[Always up to date: ]{http://www.ncnr.nist.gov}

\date {September 5, 2002 submitted to Review of Scientific Instruments}

\begin{abstract}
  We describe the design and current performance of the high-flux
  backscattering spectrometer located at the NIST Center for Neutron
  Research.  The design incorporates several state-of-the-art neutron
  optical devices to achieve the highest flux on sample possible while
  maintaining an energy resolution of less than 1\,$\mu$eV.  Foremost
  among these is a novel phase-space transformation chopper that
  significantly reduces the mismatch between the beam divergences of
  the primary and secondary parts of the instrument.  This resolves a
  long-standing problem of backscattering spectrometers, and produces
  a relative gain in neutron flux of 4.2.  A high-speed Doppler-driven
  monochromator system has been built that is capable of achieving
  energy transfers of up to $\pm 50\,\mu$eV, thereby extending the
  dynamic range of this type of spectrometer by more than a factor of
  two over that of other reactor-based backscattering instruments.
\end{abstract}


\maketitle

\section{Introduction}

Neutron scattering is an invaluable tool for studies of the structural
and dynamical properties of condensed matter.  Neutron sources produce
neutrons with wavelengths that span the interatomic spacings in solids
or the diameter of complex macromolecules, while at the same time
having energies that match, respectively, the lattice vibrational
frequencies in solids or the slow diffusive motions of atoms.  The
particular neutron scattering technique known as backscattering
\cite{Maier,Alefeld} is able to resolve energies below 1\,$\mu$eV,
which is well beyond the reach of conventional triple-axis and neutron
time-of-flight spectrometers.  Thus neutron backscattering
spectroscopy is ideally suited to the study of dynamics such as slow
motions in complex liquids, jump diffusion, and quantum rotational
tunneling.

The principle limitation of backscattering spectrometers has long been
the relatively low neutron flux on sample that they produce.  This is,
of course, a direct consequence of the excellent energy resolution
they provide.  In this paper we report on the design and performance
of the new high-flux backscattering spectrometer (HFBS) located at the
NIST Center for Neutron Research that addresses this limitation.
Compared to other backscattering spectrometers, the HFBS delivers a
higher neutron flux to the sample in large part by the use of a novel
device called a phase-space transformation chopper (Sec.~\ref{PST}).
In addition, a newly designed Doppler-driven monochromator
(Sec.~\ref{Dop}), which operates with a cam machined to produce a
triangular velocity profile, extends the dynamic range of the
spectrometer by more than a factor of two beyond that of other similar
instruments.  In a departure from other backscattering spectrometers,
the scattering chamber is operated under vacuum instead of in an argon
or helium gas environment, which improves the signal-to-background
ratio substantially.

\section{Backscattering}

\begin{figure}[t]
\includegraphics[width=2.75in]{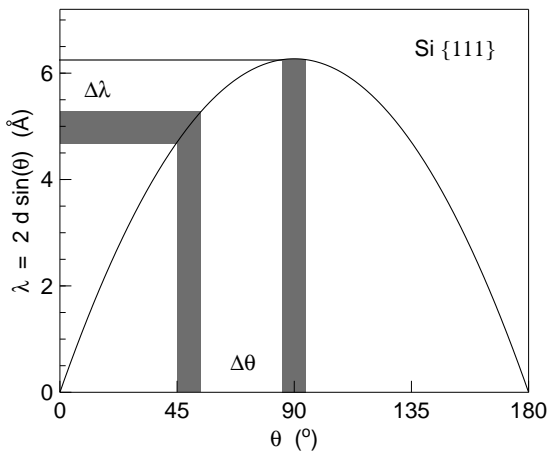}
\caption{Illustration of the backscattering principle.  Vertical
  shaded regions correspond to equal angular spreads $\Delta \theta$,
  but vastly different wavelength spreads $\Delta \lambda$ depending
  on the Bragg angle $\theta$.  As $\theta \rightarrow 90^{\circ}$,
  $\Delta \lambda$ (and thus the energy resolution) approaches a
  minimum.}
\label{backscatt}
\end{figure}

Backscattering spectroscopy exploits the fact that the wavelength
spread $\Delta\lambda$ of a Bragg-diffracted neutron beam decreases as
the scattering angle $2\theta$ approaches 180$^{\circ}$ (see
Fig.~\ref{backscatt}).  This is easily shown \cite{Bottom} by
differentiating Bragg's law ($\lambda = 2d\sin\theta$), and then
dividing the result by $\lambda$ to obtain
\begin{equation}
\label{eq:backscattering}
  \frac{\Delta\lambda}{\lambda} = \frac{\Delta d}{d} +
  \frac{\Delta\theta}{\tan\theta}\,\,.
\end{equation}
As $\theta \rightarrow 90^{\circ}$ the angular term vanishes.  This
results in a minimum value of $\Delta\lambda/\lambda$, and hence the
energy resolution, that depends on the spread $\Delta d$ and average
value $d$ of the lattice spacing between crystal Bragg planes.  In the
kinematic limit this minimum is zero.  However, dynamical scattering
theory shows that the lattice gradient term $\Delta d/d$ is non-zero,
even for perfect single crystals.  In this case, $\Delta d/d$ is given
by the Darwin width of the reflection being used to monochromate the
neutron beam.~\cite{darwin} This presents a fundamental lower bound
for the energy resolution that can be obtained via backscattering,
which depends entirely on the structure factor of the reflection being
used to monochromate the beam, and the number density of unit cells
within the monochromating material.  In the backscattering condition
the neutron beam is normal to the Bragg planes, corresponding to a
Bragg angle $\theta = 90^{\circ}$.  However, the neutron trajectories
in a beam are never perfectly parallel.  Therefore, some neutrons will
strike the crystal Bragg planes at angles slightly

%
%

\noindent
less than $90^{\circ}$, thereby satisfying the Bragg condition at
different values of $\lambda$.  Consequently the spread $\Delta
\theta$ in incident angle will also contribute to $\Delta
\lambda$.~\cite{Birr} Note that $\Delta \theta$ isn't necessarily (and
usually is not) equal to the beam divergence, as it is most often set
by the ratio of the source size to the distance between source and
Bragg planes.  If $\Delta\theta$, is small, then
\begin{equation}
\label{eq:resolution}
\frac{\Delta E}{E} = 2\frac{\Delta\lambda}{\lambda} =
2\left(\frac{\Delta d}{d} + \frac{1}{8}(\Delta\theta)^2\right),
\end{equation}
Most backscattering instruments use the \{111\} lattice planes of
perfect silicon crystals to monochromate the incident beam as well as
to analyze the energy of the scattered beam.  This is true for the
HFBS as well, so for the sake of convenience we define $\lambda_0 = 2d
= 6.2712$\,\AA,~\cite{si220} $k_0 = 2\pi/\lambda_0 =
1.00$\,\AA$^{-1}$, $v_0 = 630.8$\,m/s, and $E_0 = 2.08$\,meV.  In this
case the lattice gradient term $\Delta d/d = 1.86 \times 10^{-5}$.  As
an example of how much the angular spread $\Delta \theta $ contributes
to the energy resolution, one would need a $\Delta\theta=0.70^{\circ}$
to match the lattice gradient contribution to the energy resolution,
which is a small angular spread for a neutron beam.
Equation~(\ref{eq:resolution}) would then imply an energy resolution
of about 0.16\,$\mu$eV for the diffracted beam.

The maximum momentum transfer accessible given neutrons of wavelength
$\lambda_0$ is $Q = 4\pi/\lambda_0 = 2.00$\,\AA$^{-1}$, whereas
practical considerations generally limit the minimum useful $Q$ to
$\sim 0.1$\,\AA$^{-1}$ due to the non-zero divergence of the neutron
beam incident on the sample.  The energy range over which the
dynamical properties of a sample can be studied is set by how much the
energies of the incident and scattered neutron beams can be shifted
relative to each other.  This shift cannot be achieved by varying the
Bragg angle of the monochromator, as is often done on a triple-axis
spectrometer, because doing so ruins the excellent energy resolution.
Instead, energy transfers are obtained using other methods such as
varying the temperature of the monochromator with respect to that of
the analyzer, resulting in a continuous change in the
$d$-spacing,~\cite{Cook} or via a Doppler motion of the monochromator
crystal,~\cite{Birr} which is the method chosen for the HFBS.  Since
the analyzer crystals are fixed, a backscattering spectrometer can be
compared to a triple-axis instrument operating in a fixed final energy
configuration.  This is also often referred to as an inverted geometry
configuration.  Typical backscattering instruments with sub-$\mu$eV
resolution can reach energy transfers from $\pm 10$ to $\pm
15$\,$\mu$eV using earlier style Doppler-driven monochromator
systems.~\cite{illyb}

\section{General Spectrometer Layout}

%
%
\begin{figure}[b]
\includegraphics[width=3.2in]{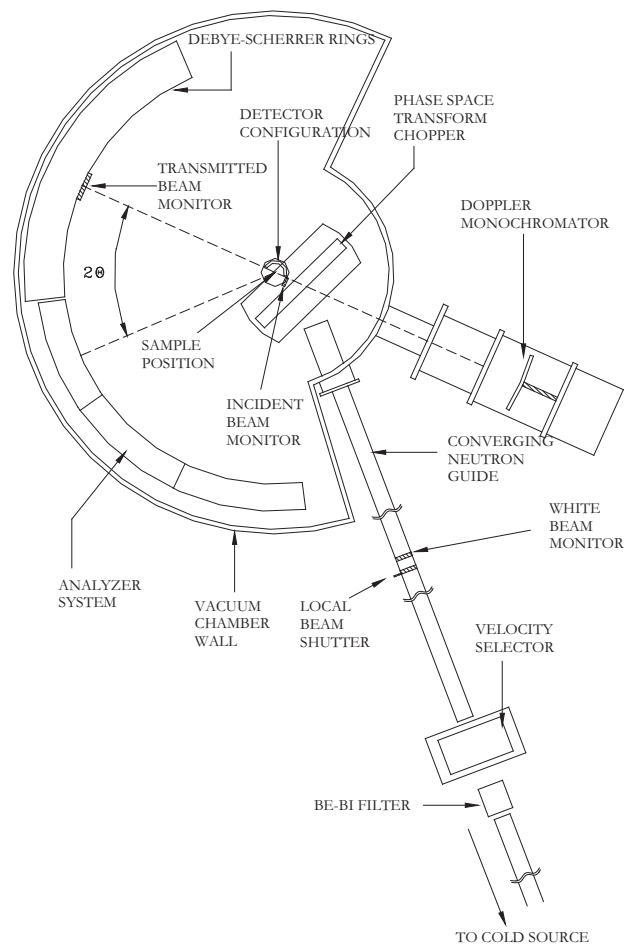}
\caption{ General layout of the NIST Center for Neutron Research
  high-flux backscattering spectrometer (HFBS).}
\label{layout}
\end{figure}

The design of the HFBS backscattering spectrometer is optimized to
provide a large dynamic range and the highest neutron flux on sample
possible while maintaining a sub-$\mu$eV energy
resolution.~\cite{GeNe98} To achieve these goals the HFBS design
incorporates several state-of-the-art neutron optic devices, which are
identified in the schematic diagram of the spectrometer shown in
Fig.~\ref{layout}.  Neutrons from the cold source of the 20\,MW NCNR
research reactor (Sec.~\ref{Source}) are conducted along a 41.1\,m
straight neutron guide that is 15\,cm high by 6\,cm wide, and pass
through beryllium and bismuth filters and a velocity selector.  A
converging guide (Sec.~\ref{CG}), located after the local beam
shutter, focuses the neutron beam cross section down to 2.8\,cm
$\times$ 2.8\,cm, which enhances the neutron flux by $\simeq 3.9$.
The neutrons then encounter a phase space transformation (PST) chopper
(Sec.~\ref{PST}), a device that Doppler-shifts the incident neutron
wavelength distribution towards the desired backscattered wavelength
$\lambda_0$.  The PST chopper provides an additional gain of 4.2 in
neutron flux, but at the expense of a sizable increase in divergence.
However in doing so the PST chopper alleviates the severe divergence
mismatch between the primary and secondary spectrometers, which has
been a long-standing problem with backscattering instruments.

Continuing on from the PST chopper, the neutrons are backscattered
from a spherically focusing monochromator, strike the sample, and then
are backscattered a second time from a spherically focusing analyzer
system before they finally reach the detectors.  Note that
backscattered neutrons must pass through the sample twice.  Both the
monochromator (Sec.~\ref{Mono}) and analyzer (Sec.~\ref{Ana}) are
composed of large, bent, silicon \{111\} crystals.  These crystals are
intentionally bent to increase the lattice gradient term in the energy
resolution (see Eq.~(\ref{eq:resolution})) in order to obtain a
roughly three-fold increase in neutron count rate, while maintaining
the sub-$\mu$eV energy resolution constraint.  Finally, a large energy
difference (dynamic range) of up to $\pm$50\,$\mu$eV can be
established between the monochromated and analyzed beams using a
cam-based Doppler drive that produces an oscillatory motion of the
monochromator (Sec.~\ref{Dop}).

The excellent energy resolution of backscattering instruments comes at
the cost of an inherently low neutron flux on the sample.  But gains
provided by the converging guide, the PST chopper, and the strained
silicon monochromator and analyzer crystals help to compensate for
this.  In addition, the HFBS analyzer stands 2\,m tall, spans
165$^{\circ}$ in $2\theta$, and subtends nearly 23\,\%\ of $4\pi$
steradians (Sec.~\ref{Ana}), making it the largest analyzer of any
other backscattering instrument.  It is composed of about 12\,m$^2$ of
silicon.  This large analyzer provides yet another gain over other
spectrometers through an increased count rate that is not included by
measurements of the neutron flux on the sample.  In the following
subsections we discuss the various components of the spectrometer in
detail starting from the cold source and ending with the detectors.
In Sec.~\ref{Perform} we report on the performance of the HFBS and
present results from several experiments.  Recent upgrades and some
potential prospects for improvements to the spectrometer are discussed
last in Sec.~\ref{prospects}.

\subsection{Neutron Source and Straight Guide}
\label{Source}

The HFBS is located at the end of a dedicated straight neutron guide,
labelled NG-2, that directly views a liquid hydrogen cold source.  The
cold source exhibits a Maxwellian spectrum with an effective
temperature of 37\,K for neutrons with wavelengths 4\,\AA\ $\le \lambda
\le$ 15\,\AA.~\cite{Williams} The flux at the backscattered wavelength
$\lambda_0$ is estimated to be $\sim 1.2 \times
10^{11}$\,n\,(cm$^2$\,s\,Sr\,\AA)$^{-1}$. Neutrons from the cold
source reach the HFBS by traveling along an evacuated glass guide
41.1\,m in length with a constant 15\,cm high by 6\,cm wide cross
section.  The interior top and bottom surfaces of the guide are coated
with NiCTi supermirrors, which have a critical angle for reflection
given by $\theta_c = Q_c\lambda = 0.044 \lambda$\,\AA$^{-1}$.  The
interior side surfaces of the guide are coated with
$^{58}$Ni-equivalent supermirrors for which $\theta_c = 0.026
\lambda$\,\AA$^{-1}$.~\cite{Chuck}


An 87\,cm long gap interrupts the straight guide 26.3\,m
downstream from the cold source to provide space for filter
material and a velocity selector.  These elements are necessary
because the HFBS design places the sample position and detectors
in close proximity to a direct line of sight with the reactor
core.  The estimated loss of $\lambda_0$ neutrons from this gap is
$\approx 15$\%.  Three blocks of vacuum-cast polycrystalline
beryllium and one block of ``pseudo'' single-crystal bismuth, each
10\,cm in length and 16.5\,cm $\times$ 6.4\,cm in cross section,
are used to remove fast neutrons and suppress core gamma-ray
($\gamma$-ray) radiation.~\cite{deGraaf} Vacuum-cast beryllium is
known to produce half the beam broadening for a given length of
filter at a given wavelength than does the hot-pressed grade of
beryllium.~\cite{Glinka} This implies fewer neutrons will be lost
through the guide walls downstream after they have passed through
the filter.  The filters are also cooled within a liquid nitrogen
dewar to minimize the effects of thermal diffuse scattering and
maximize the transmission of cold neutrons.  Using this 40\,cm
combination of beryllium and bismuth filter material, the
respective transmission of core $\gamma$-rays ($E \ge 2$\,MeV),
and fast ($E \ge 2$\,MeV), epithermal ($E \sim 1$\,eV), and cold
neutrons are estimated to be 0.20\%, 0.06\%, $2 \times 10^{-9}$\%,
and 53\%.  (The estimate for the cold neutron transmission
includes the gap loss of 15\%.)

The neutron velocity selector, placed just after the two filters,
is used to limit $\Delta\lambda/\lambda$ to 18\%.  This minimizes
the background produced by neutrons having wavelengths that lie
outside the bandwidth accepted by the PST chopper.  The blades of
the velocity selector have an axial length of 30\,cm, and are
composed of 0.4\,mm thick carbon-fiber in epoxy loaded with
$^{10}$B as absorber material. This gives a relative suppression
of unwanted neutrons of $2 \times 10^{-4}$.  The velocity selector
was designed to accept the entire 15\,cm $\times$ 6\,cm neutron
guide cross-section.  This posed a technical challenge because the
blades had to have a large enough diameter to accommodate the
15\,cm vertical dimension of the guide, yet be able to move at
high speeds.  To achieve a peak neutron transmission centered at
$\lambda_0$, the velocity selector must rotate at 16200\,rpm,
which corresponds to a tangential speed of 410\,m/s at the edge of
the blades.  The peak transmission of the selector at this
wavelength is 83\%.

A standard gold foil activation analysis was performed along the key
points of the guide with the NIST reactor operating at full power (20
MW).  This analysis assumes a cross section of 98.65 barns for thermal
neutrons with wavelength $\lambda = 1.8$\,\AA.  Just upstream of the
filters, the thermal capture flux equivalent was measured to be $3.37
\times 10^9$cm$^2$/sec.~\cite{update}  After the filters and the
velocity selector, the thermal capture flux is $1.57 \times 10^9$
cm$^2$/sec,~\cite{update} which corresponds to a net transmission of
about 44\%.  This value is consistent with the losses estimated for
the guide cut and the filter and velocity selector transmissions.

\subsection{Converging Guide}
\label{CG}

%
%
\begin{figure}[b]
\includegraphics[width=2.75in]{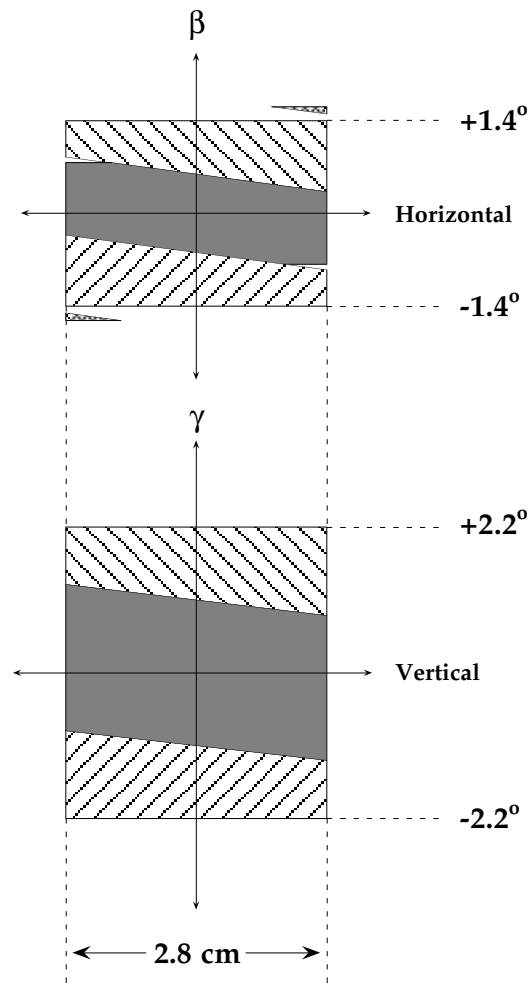}
\caption{ Acceptance diagrams for $\lambda_0$ neutrons exiting the
  converging guide.  Top panel: horizontal divergence $\beta$ versus
  lateral guide dimension.  Bottom panel: vertical divergence $\gamma$
  versus vertical guide dimension.}
\label{accept}
\end{figure}

A converging guide is used to focus the large guide beam cross section
down to a size that is commensurate with typical sample dimensions.
The guide entrance is located 41.3\,m downstream from the cold source,
just after the local beam shutter, and before the PST chopper.  All
four of its interior surfaces are coated with $2\theta^{\rm
  Ni}_c$-equivalent ($Q_c = 0.044$\,\AA$^{-1}$) supermirrors, which
compress the beam cross section from 15\,cm $\times$ 6\,cm down to
2.8\,cm $\times$ 2.8\,cm.  Acceptance diagrams \cite{Anderson,Copley}
were used to determine the optimal guide length and distances over
which the vertical and horizontal focusing should occur.  In this
context ``optimal'' does not necessarily imply the largest gain factor
(although the resulting gain is still quite good).  Instead, our
design goal was to maximize the gain subject to the constraint that
the resulting phase space elements corresponding to different
reflections remain ``compact,'' or bunched tightly together.  In so
doing one avoids the presence of gaps in the divergence of the
subsequent neutron beam seen by the PST chopper.

Based on these considerations, the vertical focusing was designed to
take place over a 4\,m guide length, while the focusing in the
horizontal direction occurs over only the last 3\,m.  The resulting
acceptance diagrams are shown in Fig.~\ref{accept} for $\lambda =
\lambda_0$ assuming a uniform illumination of the guide entrance, and
a perfect reflectivity for the supermirror coatings for $0 \le Q \le
Q_c$.  Elements that correspond to zero and one reflection from the
converging guide walls are represented by the solid and hatched
regions, respectively.  The tiny, isolated wedge-shaped elements that
appear in the horizontal case arise from neutrons that undergo two
reflections.  These acceptance diagrams were calculated analytically.
As a cross check, Monte Carlo simulations of the straight guide $+$
converging guide system were performed, and these yielded acceptance
diagrams that agreed extremely well with those shown in
Fig.~\ref{accept}.

The performance of the converging guide is characterized by the
uniformity of the beam profile exiting the guide, and by the flux
gain.  To quantify these two parameters, autoradiograph images of the
beam before and after guide were obtained by irradiating
(independently) two separate $\sim$125\,$\mu$m thick Dy foils
(1\,$\mu$m = 10$^{-6}$\,m).  The beam intensity was then integrated
over 9 (5) different circular regions on the autoradiograph taken
before (after) the guide, with the regions in each case representing
slightly less than 2\%\ of the total beam cross sectional area.  These
intensities vary by 7\%\ or less, indicating a highly uniform beam
profile at both guide positions.~\cite{GeNe98} The measurements also
indicate a flux gain of 3.43, somewhat higher than the value of 3.1
predicted by the acceptance diagram calculations.  By contrast, a
larger flux gain of 3.33 is predicted by the Monte Carlo simulations
which take into account neutrons lost in the 87\,cm guide cut which
otherwise would have entered the converging guide, but would have been
lost in the walls.

Gold foil activation measurements were made at the entrance and exit
of the guide to obtain a more accurate value of the flux gain, and the
corresponding thermal neutron capture fluxes were $2.13 \times
10^8$\,n\,(cm$^2$\,sec)$^{-1}$ and $8.29 \times
10^8$\,n\,(cm$^2$\,sec)$^{-1}$, respectively.~\cite{GeNe98,update}
Both values are accurate to within 1\%\ assuming an uncertainty of one
standard deviation based on counting statistics.  The resulting flux
gain is then 3.89, higher than both the calculated value and that
obtained from the autoradiograph images.  This discrepancy motivated,
in part, a subsequent check of the reflectivity of the straight guide
coatings, where additional superstructure in the reflectivity of the
$^{58}$Ni-equivalent supermirror coatings was found beyond the
presumed critical wave vector of $Q_c = 0.026$\,\AA$^{-1}$.
Incorporating the results of the reflectivity studies on the guide
supermirror coatings into the simulations was then sufficient to
reconcile the discrepancy between the experimental and calculated
gains entirely.~\cite{Jeremy}

%
%
\begin{figure}[b]
\includegraphics[width=2.75in]{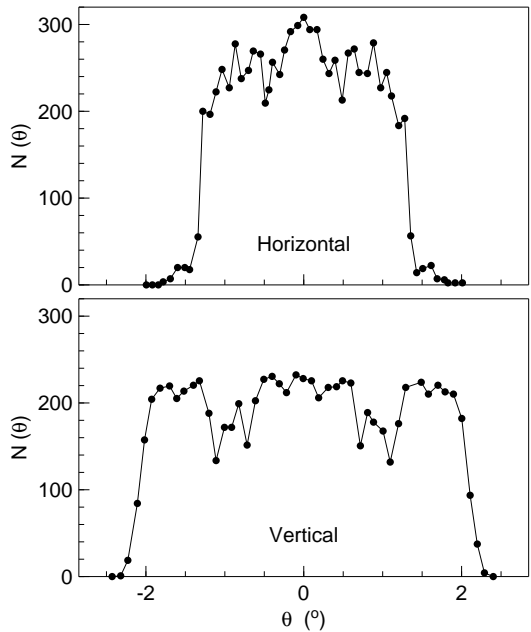}
\caption{ Monte Carlo simulations of the angular distributions of
  neutrons exiting the converging guide in the horizontal and vertical
  directions. (Simulations are courtesy of J. C. Cook.)}
\label{cook}
\end{figure}

The divergence of the neutron beam exiting the converging guide has
not been measured experimentally.  However, Monte Carlo simulations of
the angular distributions of neutrons exiting the converging guide
were done that took into account the non-zero bandwidth transmitted by
the velocity selector.  Figure~\ref{cook} shows the results of these
simulations, where $N(\theta)$ is the total number of neutrons exiting
the guide with a horizontal (vertical) divergence $\theta$ with
respect to the beam axis.  The vertical axes have been scaled such
that the two integrated distributions are equal (i.\ e.\ contain the
same number of neutrons), but otherwise the units are arbitrary.  The
distributions at the exit of the guide are found to be reasonably
uniform in both the horizontal and vertical directions.

\subsection{Phase Space Transformation Chopper}
\label{PST}

Backscattering spectroscopy is inherently a flux-limited technique
because of the narrow energy resolution it provides.  To help boost
the low count rates, most spectrometers employ a rather poor
$Q$-resolution that is introduced by the focusing analyzer system,
which compresses large regions of solid angle into few detectors.
However, all cold neutron backscattering instruments at steady state
sources are located on neutron guides, the coatings of which limit the
divergence of the beam fed to the monochromator system.  This then
creates a situation that effectively defeats the purpose of the
focusing analyzer in that the monochromator system cannot supply a
beam with sufficient divergence to take full advantage of the large
angular acceptance of the analyzer system.  Thus the opportunity
exists to increase backscattering count rates, without adversely
affecting the performance of the instrument, if one can design a
primary spectrometer capable of generating a highly divergent beam
from that supplied by the guide.  To accomplish this, Schelten and
Alefeld proposed the idea of neutron phase space transformation (PST)
using moving mosaic crystals.~\cite{Schelten}

\begin{figure}[t]
\includegraphics[width=2.75in]{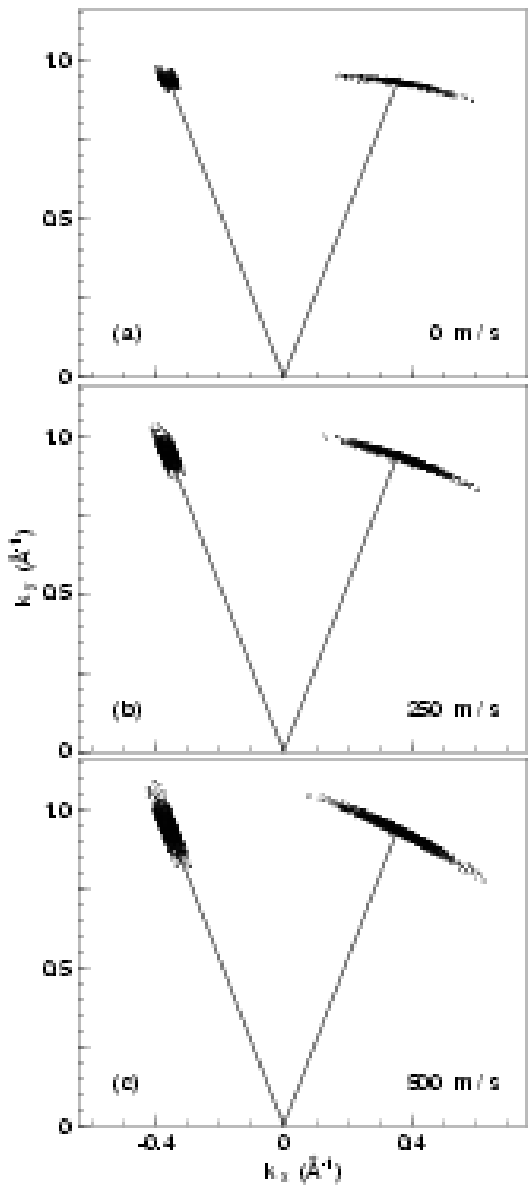}
\caption{ Simulations of the phase space transformation process using
  0.5\,cm thick HOPG crystals with an intrinsic mosaic spread of
  5$^{\circ}$ FWHM.  The different panels correspond to tangential
  crystal speeds of (a) 0\,m/s, (b) 250\,m/s, and (c) 500\,m/s.  The
  simulations are three-dimensional projections onto the $(k_x,k_y)$
  plane.}
\label{pstphase}
\end{figure}

The phase space transformation process is outlined schematically in
Fig.~\ref{pstphase}.  Panel (a) depicts an incoming, well-collimated
``white'' beam, similar to that which exits a neutron guide, as an
element in phase space.  Each point in this element corresponds to a
neutron with an incident wave vector $k_i$ measured relative to the
origin.  After diffracting from a stationary mosaic crystal, the phase
space element is transformed into the concave-shaped element shown on
the right.  No neutron energies are changed in this process, so there
is no change in the number of neutrons in a given wavelength band.
The situation changes, however, if the crystal is set in motion along
a direction perpendicular to the average scattering vector, and
antiparallel to the projection of $k_i$ onto the crystal Bragg planes
as shown in panel (b).  In this case the concave element rotates in
phase space, and this {\it does} change the neutron energy
distribution.  More importantly, the rotation is such that shorter
wave vectors become elongated while longer wave vectors are shortened,
thereby ``bunching'' up the wave vectors of the diffracted neutrons
about the desired backscattered value $k_0 = 1.00$\,\AA$^{-1}$.  Thus
the number of neutrons in a given wavelength band can be increased via
phase space transformation.  It is possible to carry this process too
far by moving the mosaic crystals too quickly.  In this case the
outgoing phase space element rotates too much, and the desired
``bunching'' effect reverses (Fig.~\ref{pstphase}(c)).

%
%

It can be difficult to understand how the phase space transformation
process works.  An intuitive explanation can be given in real space
keeping in mind the mosaic nature of the crystalline Bragg planes, as
well as the tight divergence of the incident beam.  The slower moving
neutrons incident on the PST chopper HOPG crystals must find mosaic
blocks oriented at higher angles in order to satisfy the Bragg
condition.  Therefore they get a ``push'' from the moving crystal.  By
contrast, the faster neutrons satisfy the Bragg condition at smaller
angles, so diffraction occurs from crystallites oriented in the
opposite sense, thereby reducing their speed.  The net effect of this
process is the remarkable conversion of the incoming spread in neutron
wave vector (i.\ e.\ energy) into a corresponding outgoing angular
spread, precisely the requirement for increasing the count rate of a
backscattering spectrometer.  This transformation process is
necessarily consistent with Liouville's theorem which requires that
the volumes of the incoming and outgoing phase space elements be
identical.

\begin{figure}[b]
\includegraphics[width=2.75in]{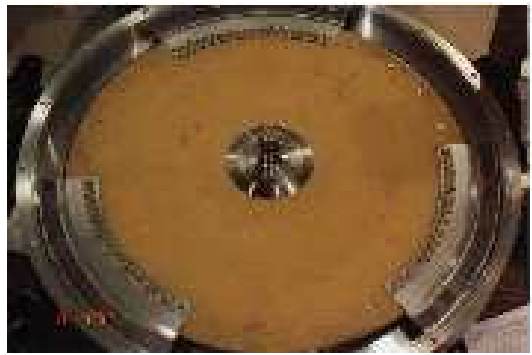}
\caption{ Photograph of the PST chopper with the front casing removed.
  The chopper disk and the three cassettes which contain the HOPG
  crystals are exposed.  The distance between the axis of rotation and
  the center of the crystals is 505\,mm.}
\label{chopper}
\end{figure}

Because the HFBS monochromates and energy-analyzes neutrons using the
(111) reflection of Si, the PST chopper has been designed to enhance
the neutron flux at the corresponding backscattered energy $E_0$.  As
implemented on the HFBS, the device consists of a 1\,m diameter disk
whose periphery is divided into six sectors.  Alternate sectors are
covered with crystals of highly-oriented pyrolytic graphite (HOPG)
that are 34.5\,mm tall, 1.5\,mm thick, and mounted inside protective
cassettes.  These cassettes clamp the crystals firmly in place to
prevent any movement which would certainly damage the crystals given
the high speeds they experience.  Finally, the chopper disk is mounted
such that the axis of rotation is parallel to the average graphite
scattering vector.  Figure~\ref{chopper} is a photograph of the PST
chopper inside its casing that shows the three cassettes which contain
and protect the 180 HOPG crystals.  The design of the PST chopper
obviously introduces a pulsed-structure to the diffracted neutron
beam, a fact that is exploited to good advantage by the data
acquisition system described in Sec.~\ref{daq}.

%
%

Graphite is an ideal choice of crystal for the PST chopper because the
$d$-spacing of the (002) planes is 3.355\,\AA, which is slightly
larger than the 3.135\,\AA\ $d$-spacing of the Si (111) reflection
used to monochromate the neutrons.  This means that neutrons that
satisfy the backscattering condition will be diffracted from the PST
chopper at a Bragg angle of 69.2$^{\circ}$, which allows for the
convenient placement of the monochromator relative to the converging
guide (see Fig.~\ref{layout}).  Moreover high-quality HOPG crystals
having the requisite mosaic are readily available, and the HOPG
neutron reflectivity is quite good.~\cite{Shapiro} Based on results
from the analytical calculations and Monte Carlo simulations reported
in Appendix~\ref{pstapp}, the effective FWHM mosaic of the graphite
was chosen to be 7.5$^{\circ}$.  This mosaic was obtained by
sandwiching three crystals, each having a mosaic between
2.25$^{\circ}$ and 3.00$^{\circ}$, between wedge-shaped spacers.  This
approach has the advantage of limiting the vertical mosaic to that of
an individual crystal, or about 2.5$^{\circ}$, and hence the vertical
divergence.  This is an important consideration from a design
standpoint for two reasons.  First, the vertical divergence determines
the height of the monochromator.  Second, given the high accelerations
reached by the Doppler drive (see Sec.~\ref{Dop}), the mass of the
monochromator must be kept a low as possible, and a larger height will
imply a larger mass.

%
%
\begin{figure}[b]
\includegraphics[width=2.75in]{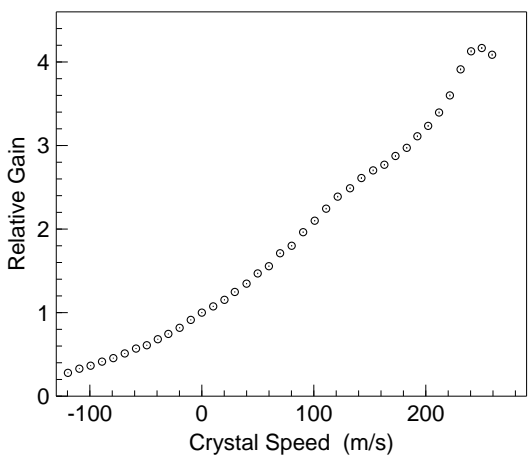}
\caption{ Relative flux gain as a function of the tangential speed of
  the HOPG crystals in the PST.  At the operating speed of 250\,m/s
  (4730\,rpm), the PST chopper gives a maximal increase in neutron
  flux of 4.2 compared to the chopper at rest.}
\label{pstgain}
\end{figure}

%
%
\begin{figure}[t]
\includegraphics[width=2.75in]{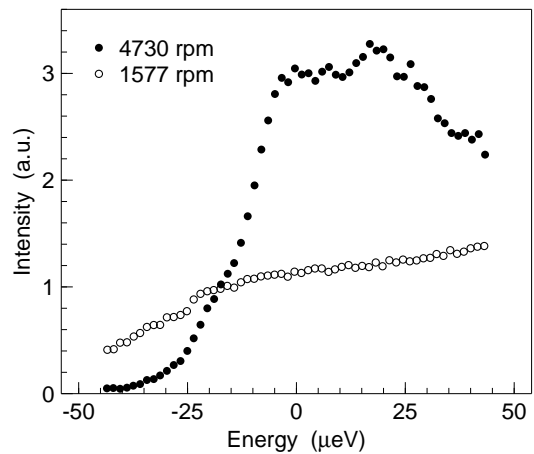}
\caption{ Neutron beam intensity as a function of energy transfer,
  measured with the incident beam monitor for two different
 operating speeds of the PST.
}
\label{monitor}
\end{figure}

A linear crystal speed of 250\,m/s was chosen, again based on the
calculations described in Appendix~\ref{pstapp}.  At this speed the
analytical and Monte Carlo results predict gains of 6.7 and 5.0,
respectively, given the incident neutron distributions produced by the
guide system described earlier.  To achieve a speed of 250\,m/s, the
PST chopper must rotate at 4,730\,rpm. Note that the choice of chopper
radius (0.505\,m), rotational frequency (79\,Hz), and number of
cassettes all depend on the speed of $\lambda_0$ neutrons (630.8\,m/s)
and the nominal distance between the monochromator and the PST chopper
(2\,m).  This is because of the critical timing issue whereby the HOPG
crystals must rotate fully out of the monochromated beam path by the
time the neutron pulse diffracted from the PST chopper and then
backscattered from the monochromator returns to the PST.  With the
aforementioned parameters, the PST chopper rotates by approximately
180$^{\circ}$ in this time, thereby allowing the now monochromatic
neutron beam to pass through an open segment in the chopper on its way
to the sample.

Because of this timing constraint, the PST chopper fully transmits the
monochromated neutron pulses at several discrete frequencies only.  In
addition to 4,730\,rpm, the HFBS can function with the PST chopper
rotating at 1/3rd speed, or 1,577\,rpm.  To avoid these timing issues
during the initial performance tests of the PST chopper, measurements
of the relative neutron flux as a function of crystal speed were made
by placing a detector directly on top of the PST chopper casing, i.\
e.\ slightly out of the backscattering condition.  These measurements
were performed in the fall of 1997, and are shown
in Fig.~\ref{pstgain}.  The measured gain at the operating velocity of
250\,m/s is 4.2, somewhat reduced from the calculations, but still
quite substantial.  The HFBS is the first neutron spectrometer to
incorporate a PST chopper.

A very important finding of the PST chopper simulations is that the
diffracted energy bandwidth has a FWHM of roughly 80\,$\mu$eV.  Thus
large energy transfers $\Delta E$ will come at the price of a reduced
flux on the sample, particularly at the extremes of the dynamic range.
Because the simulations show that the distribution of energies
reflected from the PST chopper is somewhat skewed to energies greater
than $E_0$ = 2.08\,meV, this reduction should be more pronounced for
incident energies less than $E_0$.  Fig.~\ref{monitor} displays the
flux as a function of energy transfer $\Delta E = E_i - E_f$ as
recorded by the incident beam monitor which is mounted between the PST
chopper and the sample.  A similar spectrum is collected for every
data file and is used to normalize the scattered intensity to the
incident flux.  As expected from the simulations, one sees that the
number of neutrons incident on the sample decreases as the incident
energy deviates from 2.08\,meV.  Moreover, in qualitative agreement
with the simulation, this reduction is substantially more severe for
incident energies less than $E_0$ compared to those greater than
2.08\,meV.  Further details of the phase space transformation process
are provided in Appendix~\ref{pstapp}.

\subsection{Monochromator}
\label{Mono}

\begin{figure}[t]
\includegraphics[width=2.75in]{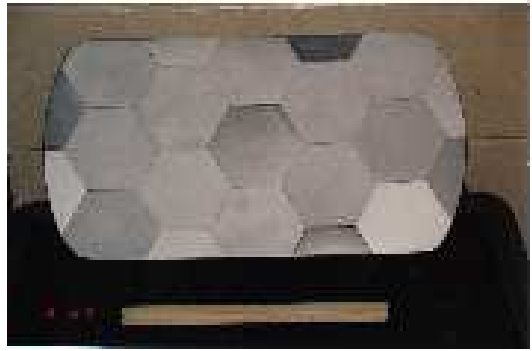}
\caption{ Photograph of the HFBS monochromator, which is composed of
  Si \{111\} wafers glued to a light-weight, but rigid, graphite
composite structure with a foam core.}
\label{mono}
\end{figure}

Figure~\ref{mono} shows a photograph of the HFBS monochromator which
is 52\,cm wide by 28\,cm tall, and spherically curved to a radius of
2.12\,m.  This radius of curvature is required to focus the neutrons
diffracted by the PST chopper (2\,m away) onto the sample which sits
2.25\,m from the monochromator.  The monochromator is composed of
$\sim 15$ hexagonal Si\,\{111\} wafers, each 750\,$\mu$m thick, glued
onto the concave surface of a support structure.  Details regarding
the choice of crystal thickness are given in Sec.~\ref{Ana} which
discusses the design of the analyzer.  The support structure is made
of a graphite composite with a foam core to minimize the total mass
(0.74\,kg without crystals, 0.95\,kg with crystals and support shaft).
It is designed to deflect no more than 0.25\,mm at the highest Doppler
drive speed because large deflections would broaden the energy
resolution.  The two-component epoxy used to attach the silicon wafers
to the support was selected for its durability under dynamic loading
conditions.  The wafers were originally glued onto the monochromator
support using a single-component anaerobic retaining adhesive widely
used in the automotive industry.  However, during the initial
commissioning phase of the spectrometer there were numerous adhesive
failures, particularly at high monochromator speeds.  These resulted
in the catastrophic loss of many wafers and prompted the search for an
alternative.  The two-component epoxy has proven to be very reliable,
and there have been no failures since its first use.

A large monochromator width is needed to span most of the neutron beam
diffracted from the PST chopper which, as a consequence of the phase
space transformation process, has a much larger horizontal divergence
($\sim 15^{\circ}$) compared to that of the beam exiting the
converging guide ($\sim 2.8^{\circ}$).  The width of the monochromator
was specifically chosen so that it would capture the FWHM of the
neutron beam diffracted from the PST chopper.  Assuming a Gaussian
angular distribution for the diffracted beam, this width is sufficient
to intercept nearly 80\%\ of the neutrons arriving at the
monochromator position.  By comparison, to intercept 95\%\ of the beam
would require a monochromator width of more than 85\,cm, which is
impractical due to the mass limitations imposed by the Doppler drive.

%
%

Measurements were performed to verify that most of the neutrons
arriving from the PST chopper do indeed strike the monochromator.
These were done by sequentially masking all but the left, middle,
and right thirds of the monochromator surface with a neutron
absorbing material, and then measuring the neutron intensity
scattered by a standard vanadium sample with the Doppler drive
operating at low speed.  These measurements indicate that more
than 40\%\ of the monochromatic neutrons that scatter from the
sample come from center third of the monochromator.  This is
consistent with the value of 40\%\ expected for a Gaussian
distribution truncated at half-maximum. Measurements of the
neutron intensity produced by the left and right thirds of the
monochromator are symmetric, and also consistent with what is
expected for a Gaussian distribution ($\sim$ 30\%\ on either
side).

Additional measurements were made with the monochromator divided
vertically into thirds.  These results indicate that roughly 50\%\
of the neutrons that strike the sample come from the center third
of the monochromator, while approximately 25\%\ come from each of
the thirds above and below the center.  If one assumes a vertical
Gaussian distribution (which is consistent with these results),
then these measurements suggest that the monochromator is tall
enough to accept more than 95\%\ of the neutrons leaving the PST
chopper.  When taken together, the vertical and horizontal
measurements indicate that the monochromator is large enough to
accept $\sim$ 75\%\ of the neutrons that leave the PST chopper. In
fact, since the simulations described in Appendix A indicate that
the angular distribution of neutrons is truncated, we believe that
the fraction of neutrons that impinge on the monochromator is
actually somewhat greater than 75\%.

\subsection{Doppler Drive}
\label{Dop}


The HFBS employs a mechanical Doppler drive to produce an oscillatory
motion of the monochromator that is used to doppler shift the neutron
energies incident on the sample about the average value $E_0$.  The
motion of the monochromator is oriented along the average silicon
wafer [111] direction.  In this manner the backscattering condition is
always maintained, thereby preserving the desired sub-$\mu$eV energy
resolution.  Many authors have discussed Bragg diffraction from moving
lattices in detail.~\cite{Shull,Stoica} Backscattering from a moving
monochromator is a special case of this for which the neutron velocity
is parallel to the motion of the crystal Bragg planes.  In this case,
the energy shift of backscattered neutrons relative to $E_0$ is given
by
\begin{equation}
  \Delta E = E_m - E_0 = 2E_0\left(\frac{v_m}{v_0}\right)+
  E_0\left(\frac{v_m}{v_0}\right)^2\, ,
\end{equation}
where $v_m$ is the velocity of the monochromator.

%
%
\begin{figure}[t]
\includegraphics[width=2.75in]{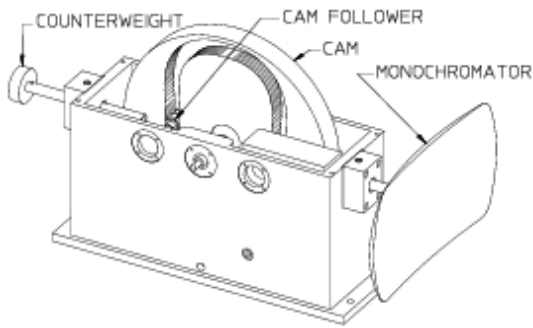}
\caption{Internal schematic view of monochromator doppler drive and
  cam system.  From Gehring and Neumann.~\protect\cite{GeNe98}}
\label{doppler}
\end{figure}

In contrast to other backscattering spectrometers, the HFBS employs a
cam-based Doppler drive system to vary the incident energy $E_i$ (see
Fig.~\ref{doppler}).  The velocity profile is determined uniquely by
the shape of the cam, which is machined from tool steel and then heat
treated.  The monochromator is connected through a shaft to bearings
that follow the contours of the cam, thereby converting the rotational
motion of the cam into an oscillatory linear motion.  A counterweight
of equal mass is driven in opposition by the same method to maintain
the dynamical balance of the Doppler drive system.  The dynamic range
of the spectrometer is set by the frequency of the Doppler drive for a
given cam.  Cams with triangular and sinusoidal velocity profiles have
been tested on the HFBS.  Both have certain advantages over the other.
A triangular velocity profile is desirable because the linear portions
weight all energy transfers equally in time.  A sinusoidal waveform,
on the other hand, spends more time at the maximum speeds, thereby
weighting higher energy transfers more heavily.  This can be useful
when the signal is weaker at large energies.  The more commonly used
crank-shaft driven Doppler systems produce a close approximation to a
sinusoidal velocity profile.~\cite{illyb}.

\begin{figure}[t]
\includegraphics[width=2.75in]{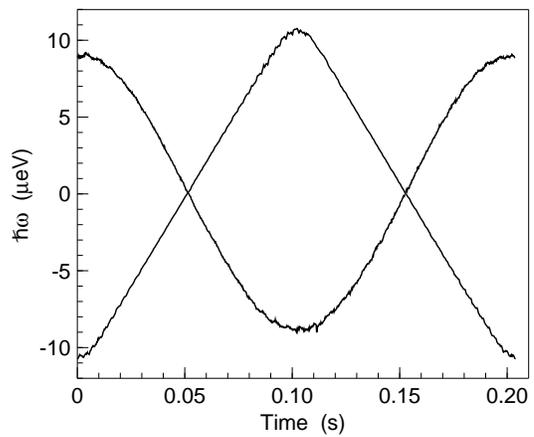}
\caption{ Velocity profiles of the monochromator during one cycle of
  the Doppler drive running at 4.94\,Hz.  The two profiles, one
  using the triangular cam, and one using the sinusoidal cam, were
  measured with a laser vibrometer.  The sinusoidal profile has been
  multiplied by -1 for ease of comparison.  Compared to the
  sinusoidal cam the triangular cam provides large regions of linear
  velocity dependence, and a larger dynamic range at a given frequency.}
\label{velpro}
\end{figure}

Figure~\ref{velpro} shows the triangular and sinusoidal velocity
profiles for each cam as measured using a laser vibrometer.  A small
polished area in the center crystal of the monochromator reflects the
light from the laser vibrometer.  The doppler-shifted light is
compared to an internal reference signal by the vibrometer which uses
the difference to generate an extremely accurate analog output that is
directly proportional to the instantaneous velocity of the
monochromator.  The profiles in Fig.~\ref{velpro} correspond to a
frequency of 4.94\,Hz, or roughly a range of $\pm 10$\,$\mu$eV in
energy transfer.  The velocity dependence of the triangular cam is
linear in time over almost the entire cycle.  In addition, the
triangular cam gives a 19\%\ larger dynamic range at the same
frequency than does the sinusoidal cam.  Interestingly, some
``ripples'' are clearly evident at higher energy transfers in both
velocity profiles.  The ripples are real, and are attributed to the
motion of the bearings in the cam that result from the non-zero
machining tolerances.  These ripples are an intrinsic feature of any
cam-based Doppler drive and can not be suppressed completely.  However
the vibrometer signal is highly localized, only reflecting the motion
of the center of the monochromator.  In fact the neutrons sample the
entire 1500\,cm$^2$ surface area.  Since the monochromator support is
not perfectly stiff, the ripples are not coherent over the surface
area. Thus the effects of the ripples are smeared out in the reflected
neutrons. As will be seen in Sec.~\ref{Perform}, they produce no
detectable effects on the energy spectra.


%
%

\begin{table}[b]
\begin{ruledtabular}
\begin{tabular}{ccc}
Monochromator\hspace*{0.5cm} &\hspace*{0.5cm}Dynamic\hspace*{0.5cm} & \hspace*{0.5cm}Energy        \\
Frequency \hspace*{0.5cm}    & Range          & \hspace*{0.5cm}Resolution    \\ \hline
4.9\,Hz\hspace*{0.5cm}       & $\pm11\,\mu$eV & \hspace*{0.5cm} 0.80\,$\mu$eV \\
9.0\,Hz \hspace*{0.5cm}      & $\pm20\,\mu$eV & \hspace*{0.5cm} 0.87\,$\mu$eV \\
16.2\,Hz \hspace*{0.5cm}     & $\pm36\,\mu$eV & \hspace*{0.5cm} 1.01\,$\mu$eV \\
\end{tabular}
\end{ruledtabular}
\caption{Instrumental elastic energy resolution as a function of
  dynamic range.}
\label{resolution}
\end{table}

The Doppler drive is designed to achieve a top frequency of 25\,Hz
corresponding to monochromator accelerations in excess of 100\,$g$.
This not only poses the problem of keeping the crystals attached to
the monochromator support, but it also implies significant vibrations
of the Doppler drive at high frequencies.  Measurements made with an
accelerometer indicate a number of pronounced resonant frequencies at
which the Doppler drive should not be operated.  The vibrations in the
Doppler drive create a distribution of velocities of the monochromator
around $E_0$, and therefore lead to a broadening of the instrumental
energy resolution.  For example, a measurement at 21\,Hz ($\Delta E =
\pm 46.7$\,$\mu$eV) gives an asymmetric instrumental energy resolution
function with a width of about 4\,$\mu$eV (FWHM).  There are several
operating frequencies at which this broadening is minimized as shown
in Table~\ref{resolution}.


\subsection{Analyzer}
\label{Ana}

\begin{figure}[b]
\includegraphics[width=2.75in]{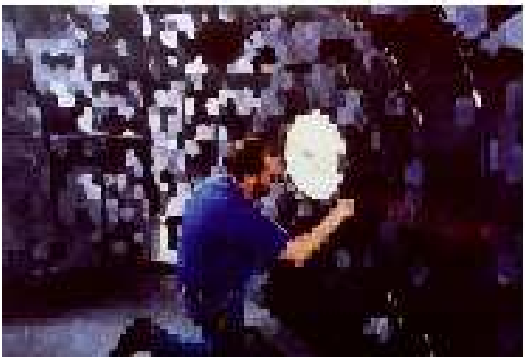}
\caption{ Photograph of the HFBS analyzer with technician S.\ Slifer
  kneeling to inspect the alignment.}
\label{analyzers}
\end{figure}

Figure~\ref{analyzers} shows a photograph of the HFBS analyzer.  The
analyzer consists of 8 spherical ``orange peel''-shaped sections that
stand 201\,cm tall by 36.4\,cm wide (at the center).  Together these
sections give an approximately continuous coverage over scattering
angles $2\theta$ ranging from 39.3$^{\circ}$ to 124.3$^{\circ}$.  For
lower angles, four Debye-Scherrer rings cover $7.8^{\circ} \le
|2\theta| \le 39^{\circ}$.  At a coverage of nearly 23\%\ of 4$\pi$
steradians, the HFBS analyzer is the larger than that of any other
backscattering instrument.  The analyzer radius of curvature is
2.05\,m, instead of 2\,m, so that backscattered neutrons are focused
onto the detectors rather than onto the sample.  This allows for a
smaller entrance window into the detector assembly, and thus a reduced
background, but at a cost of a small increase in energy resolution and
a slightly asymmetric shape to the instrumental energy resolution
function.

%
%

An important gain in intensity is obtained by gluing large diameter
($\sim$\,120\,mm) Si \{111\} wafers onto both the monochromator and
the analyzer support structures because bending increases the
intrinsic lattice gradient of silicon beyond its Darwin limit.  This
results in more flux at the expense of an increased instrumental
energy resolution.  The amount by which $\Delta d/d$ changes with
bending depends on both the radius of curvature $R_c$ and the crystal
wafer thickness $t$ according to the expression
\begin{equation}
  \frac{\Delta d}{d} = \left(\frac{\Delta d}{d}\right)_{\rm Darwin} +
  P_{\rm eff}\left(\frac{t}{R_c}\right),
\end{equation}
where $P_{\rm eff}$ is an effective Poisson's ratio which, for
spherically bent Si \{111\}, is about 0.44.~\cite{Stoica}

\begin{figure}[b]
\includegraphics[width=2.75in]{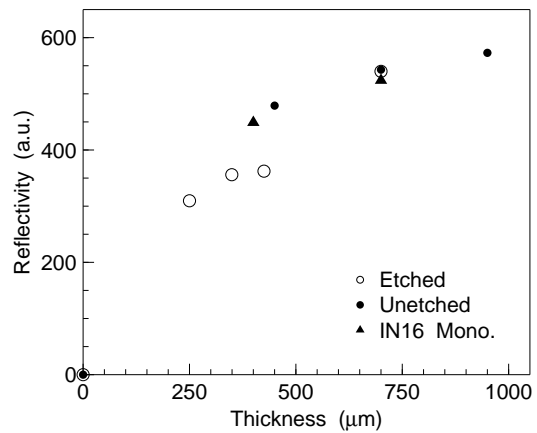}
\caption{ Plot of neutron reflectivity versus Si \{111\} wafer
  thickness.  Open circles represent etched wafers, while solid
  circles represent unetched wafers.  The solid triangles refer to the
  two IN16 monochromators.}
\label{crythick}
\end{figure}

The preceding equation predicts that a thickness of only 150\,$\mu$m
is sufficient to obtain an energy resolution of 0.75\,$\mu$eV FWHM,
which satisfies the HFBS sub-$\mu$eV constraint.  However, the effect
of a significant bending strain on the reflectivity of perfect silicon
crystals was not known.  Therefore two sets of extensive tests were
carried out at the Institut Laue Langevin on the backscattering
spectrometer IN16 using small analyzer prototypes composed of wafers
with thicknesses from 250 to 950\,$\mu$m to determine the optimal
wafer thickness experimentally.  The results of these measurements
indicated a much weaker dependence of the energy resolution on the
wafer thickness $t$.  Moreover, as shown in Fig.~\ref{crythick}, it
was observed that silicon thicknesses less than 700\,$\mu$m did not
fully saturate the reflectivity.  Based on these results a value of
750\,$\mu$m was chosen for the thickness of the silicon wafers
covering both the monochromator and analyzer.  In addition, it was
found that chemically etching the wafers gave a highly Gaussian shape
to the resolution function, whereas unetched wafers resulted in an
additional strong Lorentzian component.  Hence all wafers used to make
the HFBS analyzer and monochromator were chemically etched after
cutting (but not polished).~\cite{etch}

%
%

That such thick wafers are needed to saturate the neutron reflectivity
is surprising.  The primary extinction length \cite{Bacon} for the
(111) reflection of silicon is only 34.2\,$\mu$m.  Popovici later
pointed out that for small deflections, a wafer simply bends and the
strain remains zero in the middle (neutral) layer.~\cite{Mihai} The
reflectivity will then saturate as long as this neutral layer is
thicker than the primary extinction length.  But when the deflection
exceeds roughly half the wafer thickness, which happens very soon with
thin wafers, this neutral layer becomes strained.  The wafer stretches
non-uniformly on bending and the lattice gradient changes drastically,
thereby altering the primary extinction length.

The stretching should result not just in a spread of $d$-spacings, but
also in a non-uniform change in the nominal $d$-spacing, which in
backscattering will matter.  One then has to increase the thickness in
order to return to an unstrained middle surface, and thus to a good
reflectivity.  Given $R_c$ = 2.05\,m, and an effective wafer
``diameter'' of order 102\,mm (the wafers are hexagonal), the
resulting deflection is about 0.63\,mm.  Based on this argument, the
wafer thickness should exceed 1.2\,mm in order to be larger than twice
the deflection.  Our experimental tests show that such thick wafers
are not necessary to saturate the reflectivity.  However they do
convincingly demonstrate the validity of the premise that the neutral
layers in a thin crystal wafer are drastically affected by bending.

\subsection{Detector Assembly}
\label{Deta}

Neutrons backscattered from the analyzer must pass through the
sample a second time in order to reach the detectors, which are
located on a 10\,cm radius centered on the sample position (see
Fig.~\ref{layout}). A radial collimator, composed of 3\,cm long
cadmium fins sandwiched between each detector, sits between the
sample and detectors to reduce the background and suppress
spurious scattering from the sample environment.  The HFBS
detector assembly consists of 12 12.27\,mm diameter pencil-style
detectors, each with a $^3$He fill pressure of 6\,bars.  The
detectors are located at positions corresponding to scattering
angles of 32.3$^{\circ} \le 2\theta \le 121.25^{\circ}$, or
momentum transfers of $0.56 \le Q \le 1.75$\,\AA$^{-1}$
respectively.  Each detector has an angular acceptance of
7.75$^{\circ}$, which is about half the horizontal divergence of
the neutron beam.  The PST chopper housing partially shadows the
analyzer at large scattering angles resulting in a reduced count
rate in the detectors at $Q = 1.68$\,\AA$^{-1}$ and $Q =
1.75$\,\AA$^{-1}$.

\begin{figure}[b]
\includegraphics[width=2.75in]{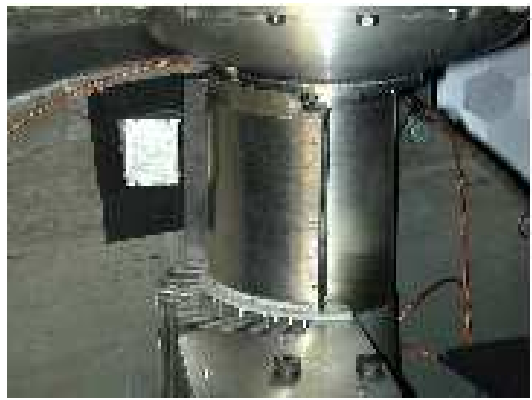}
\caption{Picture of the HFBS sample area as viewed from inside the
large vacuum chamber, which houses the PST chopper, analyzer, and
detector assembly.}
\label{sample_area}
\end{figure}

A picture of the sample area taken from within the HFBS vacuum
chamber is shown in Fig.~\ref{sample_area}.  The detector assembly
appears on the left-side of the cylindrical sample insert, and is
shielded by a cadmium and boronated-aluminum frame. The assembly
and frame are electrically isolated, and mounted onto a metal
plate attached to a stand-alone post that can be rotated away from
the sample insert to facilitate testing and repairs. The large
circular outline of the PST chopper is visible behind the sample
insert, and the chopper window through which the main guide and
monochromated beams pass is outlined in black.  The shiny
square-shaped metal piece to the left of the detector assembly is
the end of the beam stop for the main guide.  At right one can see
a few of the hexagonal silicon crystals comprising the high-angle
portion of the analyzer.  The shadowing of this portion of the
analyzer by the PST chopper is evident.

%
%

Three additional rectangular-shaped detectors with $^3$He fill
pressures of 2.5\,bar and a 38\,mm $\times$ 38\,mm active area are
mounted on the PST housing, but they are not visible in
Fig.~\ref{sample_area} because they are located behind the sample
insert. These are used to detect scattering from the smallest
three Debye-Scherrer rings, and cover momentum transfers of $0.25
\le Q \le 0.47$\,\AA$^{-1}$.  Another lone pencil-style detector
with a $^3$He fill pressure of 4\,bars and a diameter of 25.4\,mm
is used to detect neutrons that backscatter from the outer-most
Debye-Scherrer ring.  These detectors are somewhat off
backscattering, resulting in a slightly worsened energy resolution
and an asymmetric instrumental resolution function.  The larger
active area and shorter collimation also lead to a higher
background. Both the pencil detectors and the low-angle
rectangular detectors perform well in discriminating the neutron
signal from the $\gamma$ background.  The background from $\gamma$
radiation and electronic noise is negligible.

\subsection{Detector Electronics}

\begin{figure}[b]
\includegraphics[width=2.75in]{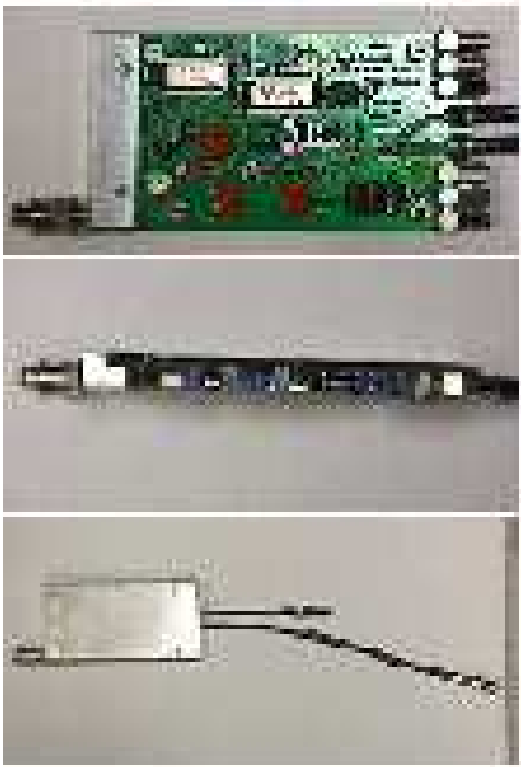}
\caption{Three views of the HFBS PAD unit:  Top - the bare circuit
board, Middle - a side view, Bottom - the complete PAD unit in its
casing.}
\label{pad_final}
\end{figure}

The HFBS uses a combined preamplifier-amplifier-discriminator
(PAD) unit to process neutron-event signals from each $^3$He
detector prior to being counted by a scaler.  Several views of the
PAD unit are shown in Fig.~\ref{pad_final}, and the manner in
which these units connect to the detector assembly can be seen in
Fig.~\ref{sample_area}.  The PAD design is non-trivial as it must
comply with a number of constraints. First, the PAD units have to
withstand operation under vacuum because the detectors are
operated inside the HFBS vacuum chamber, and the PAD units are
plugged directly onto the detectors for optimum noise reduction
and reliability. Second, the 1/2-inch (12.27\,mm) diameter pencil
detectors are physically arranged closely together to optimize
coverage.  This requires the PAD units to have a very flat form
factor.  Third, the scaler being used has ECL inputs, so the PAD
units need to produce a digital output compatible with this
standard.

%
%

Given that normal convective cooling is not present in a vacuum
environment, the first constraint was met by ensuring that the
power dissipation in all components was low enough that even the
vestigial conductive cooling through the printed circuit board,
plus whatever radiative cooling takes place, is sufficient to keep
all component operating temperatures within their limits.  Two
commercial (Amptek) hybrid ICs were chosen to perform most of the
PAD functions, the A225 preamplifier/shaping amplifier and the
A206 amplifier/discriminator.  Both ICs have milliwatt power
dissipation and are rated for vacuum operation. The digital output
buffer is based on the Telcom TC4428A dual inverting/noninverting
driver, a power CMOS part that also shows low power dissipation
even when driving heavy loads.

The required flat form factor was reached through careful
component selection and package design. The final PAD package
thickness is 11.3\,mm. This was achieved using standard
through-hole components, along with easy fabrication and assembly,
sturdiness, and adequate shielding.  Because the detectors were
mounted so closely together, the locking rings on the input SHV
plugs had to be cut off. The PAD units (and therefore the SHV
plugs which are press-fitted into them) are held in place by a pin
that is clamped into the detector mounting plate.

The TC4428A digital output driver produces an inherently balanced
TTL-level output.  The TC4428A output impedance is so low that the
necessary ECL-level outputs can be produced by the unsophisticated
but effective method of resistive voltage dividers to the -12V
supply. It is expected that future applications of this driver
would use the balanced TTL output and take advantage of the
excellent common-mode range of commercial RS485 ICs at the
receiving end.

\subsection{Operational Modes and Data Acquisition}
\label{daq}

The HFBS employs a ``real time'' data acquisition system whose primary
function is to bin the neutrons counted in each detector according to
their incident energy.  This energy is determined by the monochromator
velocity {\it at the time the neutron was backscattered from the
  monochromator}.  The data acquisition system also measures the
angular velocity and position of the doppler drive cam using a digital
encoder, and then calculates the monochromator crystal displacement
and velocity.  The velocity values are stored in a memory stack, and
then used with the displacement values to calculate the neutron flight
time to the detectors.  This flight time information is used to set
stack pointer values that the system will later need to retrieve the
correct velocity information when the neutrons are finally counted by
the scalers.  Further details of these calculations are discussed in
Appendix~\ref{ftosc}.  During each sampling period, which lasts $\sim
20$\,$\mu$s, the scaler values are latched and read.  Using previously
calculated pointer values, the incident neutron energy (i.e.\
monochromator velocity) is retrieved from the velocity stack.  The
system then increments a histogram using the appropriate incident
energy and $Q$ values.  These tasks describe the ``standard''
operational mode of the HFBS.  The data acquisition system also
supports an alternative operational mode that allows the user to
perform a ``fixed window scan'' (FWS) for which the doppler drive is
stopped and the instrument is simply measuring the elastic intensity
as a function of $Q$.  This mode is generally combined with scans of a
physical variable such as sample temperature or time.  An example of
an FWS spectrum is given in Sec.~\ref{Perform}.

%
%
\begin{figure}[b]
\includegraphics[width=2.75in]{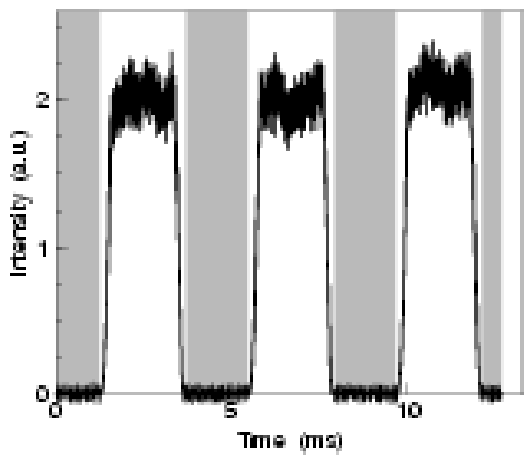}
\caption{Neutron intensity during one full cycle of the phase space
  transformation chopper with the HFBS operating in ``teach'' mode.  The
  regions of high intensity are caused by monochromated neutrons that
  scatter from the sample directly into the detectors without energy
  analysis.  Only signals that coincide with the time profile of the
  analyzed neutrons (shaded areas) are processed by the data
  acquisition system when operating in the ``standard'' mode. }
\label{dataac}
\end{figure}

A secondary function of the HFBS data acquisition system is to count
only those neutrons that backscatter from the analyzer, and to ignore
the unwanted neutrons that scatter from the sample directly into the
detectors.  This function is handled entirely in software, and the
timing is controlled by the PST chopper.  For each sampling period,
the data acquisition system measures the angular position of the PST
chopper using a two-pole resolver and then compares it to a look-up
table of values to determine if the scaler counts should be kept.  To
assist in generating this look-up table an alternative operational
mode, called ``teach'' mode, is available that bins the neutrons
counted in each detector according to the angular position of the PST
chopper.  As shown in Fig.~\ref{dataac}, the data obtained in teach
mode clearly separate the huge signal generated by the unwanted
neutrons from the much weaker signal, indicated by the shaded regions,
that results from neutrons that have been properly energy-analyzed.
Thus ``teach'' mode is used to instruct the data acquisition system
when to keep a neutron count, and when to discard it, based on the
angular position of the PST chopper.  Figure~\ref{dataac} shows one
full period of the PST chopper which, because it rotates at a constant
rate of 79\,Hz, corresponds to 12.7\,msec.  The objective of teach
mode is then to adjust the positions of the shaded regions to achieve
the highest signal-to-background ratio possible.

The HFBS data acquisition system is composed of a number of integral
hardware and software components.  The main hardware components
include a 32-bit VME-compatible single-board computer, a 32-bit latch,
a 32-bit high-speed multiscaler, an optical encoder, a
resolver-to-digital encoder, and a high-performance digital signal
processor.  The CPU provides control of the slow processes in the VME
data acquisition.  The resolver-to-digital encoder system provides a
12-bit digital output corresponding to the angular position of the PST
chopper.  The latch is used to capture and read the encoded angular
position of the PST chopper.  The scaler accumulates counts from the
neutron detectors and also counts the number of pulses sent from the
doppler drive cam optical encoder, which is used to calculate the
angular velocity and position of the cam.  A high speed digital signal
processor (DSP) is used to perform the fast VME readouts of the latch
and scaler modules, to calculate the monochromator velocity and
displacement, to calculate the pointer values, and also to increment
the histogramming memory.

\subsection{Background Reduction}
\label{backred}

The HFBS design places both sample position and detector assembly
in close proximity to the main guide beam (see Fig.~\ref{layout}).
This was done to avoid the reflectivity loss and increased beam
divergence that would result by diffracting the beam away from the
guide using a pre-monochromator.  The extra divergence is
undesirable because it would reduce the effectiveness of the
chopper.  The HFBS therefore faces the prospect of having a higher
background than those measured at other backscattering
spectrometers, where the neutron beams entering the primary
spectrometers are usually already pre-monochromated.~\cite{illyb}
Furthermore, about $6.5 \times 10^9$ neutrons per
second~\cite{update} exit the converging guide and fly towards the
PST chopper.  When this rate is compared to the single-detector
count rate of a few counts per second typical for a standard
experiment, the need for extensive background reduction is
obvious.

A number of measures have been taken to reduce the HFBS background as
much as possible.  Several of these have already been discussed, such
as the use of liquid-nitrogen-cooled beryllium and bismuth filters,
which reduce the fast neutron background component, and the velocity
selector, which limits the wavelength bandwidth seen by the PST
chopper.  The background is further reduced by continuously flushing
the flight path between the monochromator and the PST chopper with He
gas.  This process eliminates air scattering, which for dry air is
7.3\,\%\ per meter, and leads to an increase in the detector signal.
A system of masks made of absorbing boronated aluminum (Al loaded with
$\sim 5$\%\ $^{10}$B) are used to define the entrance and exit of the
PST chopper window so that neutrons exiting the converging guide are
either absorbed by the masks, or hit the rotating HOPG crystals.
Neutrons incident on the HOPG crystals are then either diffracted, or
transmitted.  To absorb those neutrons that are transmitted, a 2\,mm
thick layer of boron in epoxy, molded to the shape of the cassettes,
is positioned immediately behind the crystals.  This boron in epoxy
layer also functions as a beam stop when one of the three cassettes
has rotated into the beam path.  This is quite important because, as
seen in Fig.~\ref{dataac}, the data acquisition system is processing
signals from the detectors precisely at this time.  When none of the
crystal sections is in the beam, the incoming neutrons are absorbed
directly after the PST chopper by a beam stop made from a
$^6$Li-bearing ceramic.  An additional slit system positioned between
the chopper and sample position leads to a further reduction in
background.  Finally, the stainless steel housing of the PST chopper
is itself completely covered with boronated aluminum and an outer
layer of cadmium.

The dominant contribution to the background, in spite of the
features listed above, comes from the leakage of neutrons through
the chopper exit window during the time a crystal cassette is
blocking the beam path. We attribute this to multiple scattering
of neutrons within the chopper housing.  However, this
contribution drops by a factor of 3 when the HFBS scattering
chamber is evacuated and the chopper entrance and exit windows
(1\,mm thick aluminum) are removed, which is the standard
operating mode for the HFBS.  At the same time we observe a
22\,\%\ increase in the scattering signal.  The HFBS scattering
chamber can be evacuated with a large mechanical roughing pump to
a vacuum of better than $10^{-2}$\,mbar in less than 4 hours, at
which pressure the high voltage to the detectors can be switched
on.

\section{Performance}
\label{Perform}

A gold foil activation measurement at the sample position gives a flux
of monochromated neutrons on the sample at $E_0$ of $1.4 \times
10^{5}$\,n\,(cm$^2$\,sec)$^{-1}$.~\cite{update} The beam size at the
sample position has been measured by irradiating a 125\,$\mu$m thick
Dy foil.  The autoradiograph image is consistent with a beam size of
2.9\,cm $\times$ 2.9\,cm, and a uniform beam intensity profile.


%
%
\begin{figure}[b]
\includegraphics[width=2.75in]{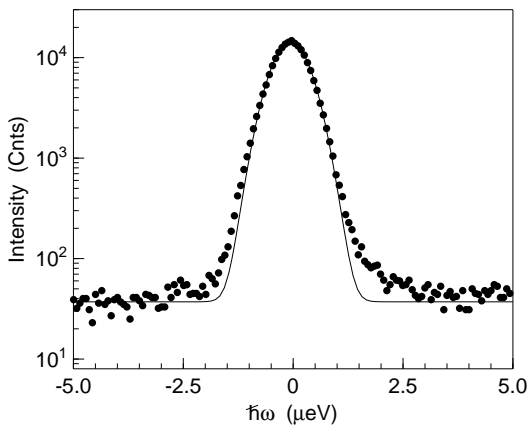}
\caption{ Standard vanadium spectrum measured on the HFBS using a
  Doppler drive frequency of 13.5\,Hz ($\pm 30$\,$\mu$eV), and
  integrating over $0.62 \le Q \le 1.6$\AA$^{-1}$).  The solid line
  represents a fit to a Gaussian function plus a constant background
  $bg\!+\!a\,\exp[-(x\,/\,b^2)]$.  The spectrum shows an almost
  Gaussian-like energy resolution with a FWHM of 0.93\,$\mu$eV.
  Extensive efforts at background reduction have led to a
  signal-to-background ratio $(a\,/\,bg)$ of better than 400:1.}
\label{van}
\end{figure}

Figure~\ref{van} shows the spectrum of an annular vanadium standard
with an outer diameter of 22\,mm, and a wall thickness of 0.88\,mm.
This geometry yields a 10\,\% scatterer.  The spectrum was taken at
room temperature with the Doppler drive operating at a frequency of
13.5\,Hz, corresponding to a dynamic range of $\pm 30$\,$\mu$eV.  The
average count rate in each detector was 106 counts per minute.  The
spectrum is shown using a logarithmic representation to show the
Gaussian nature of the vanadium spectrum more clearly.  For this
standard we observed a signal-to-background ratio of better than
400:1.  Measurements on materials having a lower cross section for
absorption than that of vanadium (e.g.\ samples with a high hydrogen
content) give signal-to-background ratios as high as 600:1.  Up to
Doppler-drive frequencies of 16.2\,Hz ($|\Delta E| \le 36$\,$\mu$eV
for the triangular cam), all HFBS spectra exhibit Gaussian-like energy
resolution lineshapes.

\begin{figure}[t]
\includegraphics[width=2.75in]{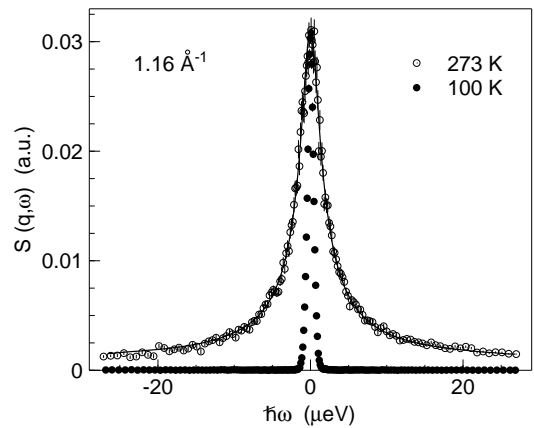}
\caption{ Dynamics of glass-forming dibutylphthalate at 1.16\,\AA$^{-1}$.
  Structural relaxation leads to a quasielastic broadening in the
  scattering law $S(q,\omega)$ with increasing temperature which can
  be described by the Fourier transform of a Kohlrausch stretched
  exponential function (solid line).  The spectrum at 100\,K --
  representing the instrumental resolution -- has been scaled down by
  a factor of 18 to match $S(q,\omega\!=\!0;273\,K)$.  }
\label{dibut}
\end{figure}

One of the main applications of neutron backscattering spectroscopy is
the study of slow atomic motions in viscous liquids.  In a first
experiment the structural relaxation in glass-forming dibutylphthalate
was measured at temperatures from 100\,K to 273\,K using a closed
cycle refrigerator.~\cite{Meyer2000} Liquid dibutylphthalate was
encapsulated inside a hollow aluminum cylinder 25\,mm in diameter,
75\,mm in length, having an annular thickness of 0.08\,mm.  This
geometry also yields a 10\,\%\ scatterer.  The spectra shown in
Fig.~\ref{dibut} have been normalized to the monitor and the vanadium
standard, and also corrected for a flat background and empty-cell
scattering.
\noindent The measuring time at 273\,K was 7\,hours at a count
rate of 76 counts per minute per detector (119 counts per minute
per detector at 100\,K).  Structural relaxation, which is
responsible for viscous flow, broadens the quasielastic linewidth
with increasing temperature. Typical for glass-forming systems is
the stretching of correlation functions over a wider time scale
than is expected for exponential relaxations.  This is an example
of a system for which the extended dynamic range of the HFBS is
especially valuable as it permits the study of structural
relaxation in viscous liquids in greater detail.

%
%

The relaxation of the Si-O network in sodium disilicate
\cite{Meyer2002} was studied at temperatures up to 1600\,K using the
HFBS high temperature furnace.  Due to the chemical reactivity of
liquid silicate glasses, a platinum sample holder had to be used.  The
neutron absorption cross section of platinum at $\lambda = \lambda_0$
is 35.9\,barns.  This is about 45 times larger than that of aluminum.
This large cross section is even more problematic given that the beam
passes through the sample twice on its way to the detectors.  Even so,
the high flux provided by the HFBS resulted in a count rate of some 13
counts per minute per detector, still enough for quantitative data
analysis.

%
%
\begin{figure}[t]
\includegraphics[width=2.75in]{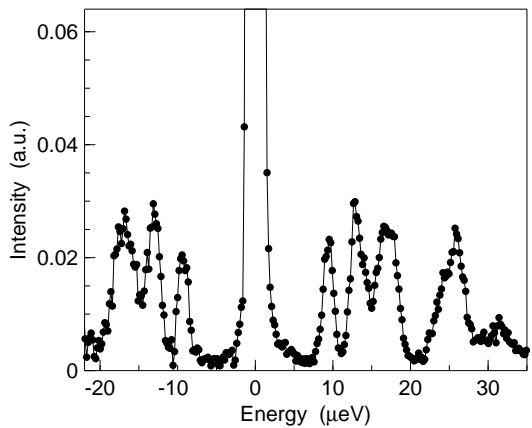}
\caption{ Tunneling spectrum of 2,6-Lutidine illustrating the large
  dynamic range of the HFBS ($\pm 36 \mu$eV), and an energy resolution
  of $1.01 \mu$eV (FWHM).  }
\label{lutidine}
\end{figure}

A recent measurement of the tunnel splitting in 2,6-Lutidine
(C$_7$H$_9$N) served as a test of the large dynamic range of the
HFBS as well as the excellent energy resolution. This system had
been examined on a number of spectrometers including a
backscattering instrument.~\cite{Fillaux} Due to the large number
of peaks in the spectrum at energies less than $50 \mu$eV, some of
which overlap, this system has also been used as a test case for
maximum entropy \cite{Mukhopadhyay} and Bayesian analysis
techniques.~\cite{Sivia2,Sivia1}  The data, summed over 10
detectors spanning 0.62\,\AA$^{-1}$ to 1.6\,\AA$^{-1}$, and
normalized to the incident beam monitor spectrum, are shown in
Fig.~\ref{lutidine}.

%
%
\begin{figure}[b]
\includegraphics[width=2.55in]{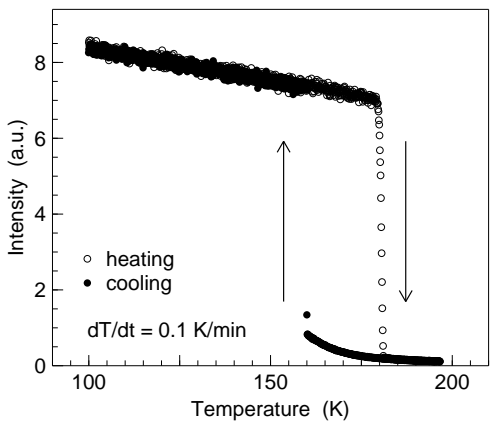}
\caption{ A fixed-window scan of toluene that clearly shows
  supercooling.  The heating and cooling rates used in this
  measurement were both $\pm0.1$ K/min.  }
\label{toluenefws}
\end{figure}

As discussed in Sec.~\ref{daq}, the backscattering spectrometer
can also be operated with the Doppler monochromator at rest, in
which case it probes only the energy-resolution-limited elastic
scattering intensity.  The FWS method is a useful technique for
determining under what conditions (e.g. temperature, time) the
dynamics of the system being studied lie within the dynamic range
of the spectrometer.~\cite{Springer77} In addition, this technique
can be useful for probing phase transitions. As an example of
this, the heating and cooling curves for toluene are presented in
Fig.~\ref{toluenefws}.  In this figure, the data were summed over
10 detectors and warming/cooling rates of $\pm0.1$\,K/min were
used.  The large hysteresis on heating and cooling, indicative of
undercooling, is clear.  The solid melts at 179\,K and the liquid
solidifies at about 160\,K.

\section{Upgrades and Prospects}
\label{prospects}

In March of 2002, work was completed on the installation of a new cold
source into the NCNR research reactor.  The complex shape of the new
source, which is an ellipsoidal shell, allows more D$_2$O to be
introduced in the cryostat chamber than did the spherical shape of the
old cold source.~\cite{Williams,newsource} This modification provides
the single biggest contribution to the overall gain in cold neutron
flux.  The center of the inner ellipsoid is also offset, so that the
thickness of liquid hydrogen nearest the source of neutrons is 30\,mm,
as compared to 20\,mm in the original source.  Finally, the inner
ellipsoid is held under vacuum, whereas the inner shell of the old
source was filled with hydrogen-vapor.  The overall flux gains range
from 40\% at 2.4\,\AA\ to slightly more than 100\% for wavelengths
greater than 15\,\AA.  For the HFBS, the new cold source has increased
the usable flux at $\lambda_0$ by a factor of 1.8, precisely in line
with expectations.~\cite{newsource} With this cold source upgrade, the
HFBS currently provides a neutron flux on sample of $2.5 \times
10^{5}$\,n\,(cm$^2$\,sec)$^{-1}$, which is higher than that of any
other backscattering spectrometer operating on a steady state source.

Several ideas for improving the HFBS have been considered and
evaluated in detail.  These include a new parabolic converging
guide~\cite{Boni} with a focal point that is centered on the PST
chopper HOPG crystals, and a supermirror guide section that
functions as an optical filter.  While a new parabolic converging
guide is mainly expected to improve the signal-to-background
ratio, an optical filter would shift the HFBS out of the direct
line-of-sight of the neutron source, and thus allow for the
removal of the beryllium and bismuth filters.  The replacement of
these filters with an optical filter could produce a substantial
enhancement of the incident neutron flux on the sample.

\section{Summary}
\label{summary}

The primary goal of the HFBS is to provide a sub-$\mu$eV energy
resolution capability to the NCNR scattering community while
maximizing the flux on sample as well as the dynamic range.  To this
end the HFBS design incorporates several state-of-the-art neutron
optic, mechanical, and electronic devices.  Foremost among these is
the phase space transformation chopper which converts the energy
spread of the neutron beam exiting the converging guide into one that
is highly divergent but more densely distributed about the
backscattered energy of 2.08\,meV.  This device has operated
flawlessly for over four years and enhances the neutron flux by a
factor of 4.2.  In addition, a considerable increase in signal is
achieved using a 4\,m long converging guide section and a large
analyzer that subtends nearly 23\,\% of $4\pi$ steradians.

In contrast to other backscattering spectrometers, the HFBS employs a
high-speed, cam-based Doppler-drive system to vary the incident energy
$E_i$ up to $\pm 50\,\mu$eV.  The shape of the cam is machined to
produce a nearly triangular velocity profile for the monochromator.
In routine operation the monochromator system provides a Gaussian-like
sub-$\mu$eV instrumental energy resolution for energy transfers of up
to $\pm$36\,$\mu$eV.  Routine user operation began in late 1999.
Since then the HFBS has proven to be a reliable instrument.  Up to now
more than 90 successful experiments, using temperatures ranging from
2\,K to 1600\,K, have been performed by users from university,
industrial, and research centers.  At present requests for beam time
on the HFBS exceed the available time by a factor of 1.8.  The HFBS
design and performance have led to sizable enhancements in both
neutron flux and dynamic range compared to that available on other
reactor-based backscattering instruments.

\section*{Acknowledgments}

The design and construction of the NIST Center for Neutron
Research HFBS spectrometer has been the result of an intense
collaboration between many scientists, engineers, technicians, and
computer programmers.  It would not be the success that it is
without the considerable expertise and efforts of the following
support staff: Technical - G.\ M.\ Baltic (head technician), A.\
Clarkson, D.\ Clem, W.\ Clow, M.\ J.\ Rinehart, and S.\ Slifer;
Engineering - P.\ Brand (Doppler drive system), C.\ Brocker (PST
chopper, vacuum chamber, and detector assembly), R.\ Christman
(Doppler drive and detector assembly), A.\ E.\ Heald (guide
shielding), J.\ LaRock (cryogenic inserts), J.\ Moyer (vacuum
chamber shielding), D.\ Pierce (velocity selector mount and
neutron shutter), I.\ G.\ Schr\"{o}der (neutron guides and
shielding); Electrical - B.\ Dickerson (PLC system), D.\ Kulp, T.\
Thai, J.\ Ziegler (detector electronics); Computer - N.\ C.\
Maliszewskyj (data acquisition sofware).  In addition, the
following scientists were involved with various aspects of the
design, construction, and testing phases of the instrument
development: Z.\ Chowdhuri, J.\ C.\ Cook (Monte Carlo
simulations), K.\ T.\ Forstner (HOPG crystal tests), P.\ D.\
Gallagher (data acquisition hardware and software), and C.\
Karmonik (PST chopper tests, background reduction, and software
development).  Special mention is due to C.\ Appel (analyzer and
monochromator) and B.\ Frick (IN16 analyzer tests), both of whom
contributed enormously towards the realization of the HFBS. The
authors finally wish to express their sincere thanks for the many
helpful and stimulating discussions with B.\ Alefeld, B.\ Frick,
A.\ Heidemann, O.\ Kirstein, A.\ Magerl, M.\ Popovici, M.\ Prager,
J.\ M.\ Rowe, and J.\ J.\ Rush.

\appendix
\section{Phase Space Transformation}
\label{pstapp}

The function of the PST chopper is to transform the shape of an
incoming neutron beam in phase space such that the neutron flux of the
outgoing diffracted beam is enhanced at the backscattering energy
$E_0$ = 2.08\,meV.  This idea was developed by Schelten and Alefeld to
address the severe mismatch in divergence between primary and
secondary spectrometers.~\cite{Schelten}

Figure~\ref{pstphase} provides a phase space diagram that outlines the
operating principle behind the phase space transformation chopper.
The incoming neutron beam has a relatively small angular divergence
that is set by the critical wave vector $Q_c$ of the guide coatings,
(Sec.~\ref{CG}) but a substantial spread in wave vector $\Delta k$ ($k
= 2\pi/\lambda$).  After diffracting from the (002) Bragg reflection
of a mosaic HOPG crystal at rest, the outgoing phase space element has
a much broader divergence (see Fig.~\ref{pstphase}(a)).
Figure~\ref{pstphase}(b) demonstrates what happens when the HOPG
crystal is set in motion parallel to $-\vec{k_x}$ at a linear speed of
250\,m/s.  The outgoing phase space element is now both larger and
more divergent.  More importantly, it is tilted such that its arc is
essentially perpendicular to $k_f$.  This tilt produces a much
narrower energy spread about $E_0$.

\begin{figure}[b]
\includegraphics[width=3.2in]{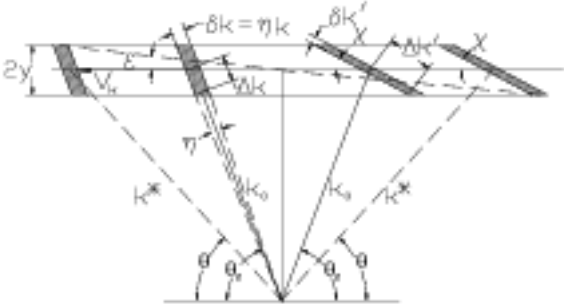}
\caption{Phase space geometry for Bragg diffraction from a moving
  crystal with mosaic $\varepsilon$ (HWHM).}
\label{pstgeometry}
\end{figure}

Analytic calculations of the gain from phase space transformation
(PST) can be made under certain simplifying assumptions, and provide
extremely useful guidance in the optimization of different physical
parameters.  The geometry for the diffraction process from the moving
lattice is shown in Fig.~\ref{pstgeometry} (reproduced from
\cite{Schelten}).  Previous calculations by Schelten and Alefeld were
performed for the gain under different crystal and spectrometer
configurations, but were done assuming the {\it optimal} crystal
velocity \cite{Schelten}.  Below we describe how the analytic
equations can be extended to arbitrary crystal velocities.

%
%

As presented in \cite{Schelten}, the angle the diffracted phase space
element makes with respect to the horizontal axis is written as
\begin{equation}
\label{eq:tanchi}
\tan\chi={{\tan{\theta}_g}\over{2\tan{\theta}_g \tan\theta+1}},
\end{equation}
where $\theta_g$ is the incident angle in the laboratory frame and
$\theta$ is the incident angle in the moving crystal frame.
$\theta_g$ is given by the $d$-spacing of the moving crystal (in this
case graphite, hence the subscript ``g'') and the desired wavelength
$\lambda_0$ via Bragg's law,
$\theta_g=\sin^{-1}\left(\lambda_0\over2d\right)$.  The velocity in
the crystal system is determined by \begin{equation}
\label{eq:vk}
V_K={{v_0}\sin{\theta_g}\lbrack{\cot{\theta_g}-\cot{\theta}}\rbrack},
\end{equation}
where $V_K$ is the velocity of the crystal and $v_0$ is the velocity
of the neutron corresponding to the desired neutron wavevector $k_0$:
\begin{equation}
\label{eq:mnvk}
{m_n}{v_0}= {{\hbar}k_0}.
\end{equation}
Equation~(\ref{eq:vk}) is obtained from the geometry in
Fig.~\ref{pstgeometry} by noting that $k_0
\sin{\theta_g}={k^*}\sin{\theta}$ and
${k^*}\cos{\theta}=K+k_0\cos{\theta_g}$, where ${\hbar}K =
{m_n}{V_K}$.

We can solve Eq.~(\ref{eq:vk}) for $\theta$ with the following result:
\begin{equation}
\label{eq:tantheta}
{\tan{\theta}}={[{\cot{\theta_g}+{{V_K}\over{{v_0}\sin{\theta_g}}}}]^{-1}}.
\end{equation}
>From Eqs.~(\ref{eq:tantheta}) and (\ref{eq:tanchi}) we can determine
the tilt angle, $\chi$, of the scattered phase space element.

>From the incident phase space element we can use the geometry
illustrated in Fig.~\ref{pstgeometry} to relate the length of the
element, $\Delta k$, to the quantities known so far.  In terms of the
mosaic spread, $\varepsilon$ (HWHM), (assumed known) we have
\begin{equation}
\label{eq:taneps}
{\tan{\varepsilon}}={{\Delta k \sin{\theta_g}}\over{K+\Delta k
    \cos{\theta_g}+k_0 \cos{\theta_g}}},
\end{equation}
which can be solved for $\Delta k$,
\begin{equation}
\label{eq:deltak}
\Delta k={{K+k_0 \cos{\theta_g}}\over{\cot\varepsilon \sin\theta_g -
    \cos{\theta_g}}}.
\end{equation}
We can also find the projection of the incident phase space element
onto the vertical axis to obtain
\begin{eqnarray}
\label{eq:h}
y & = & \Delta k \sin{\theta_g} \nonumber \\ & = & \frac{(K+k_0 \cos
  \theta_g) \sin \theta_g}{\sin \theta_g \cot \varepsilon-\cos
  \theta_g}.
\end{eqnarray}
We also note that the divergence of the incident beam, $\eta$, and the
incident beam wavevector, $k_0$, determine the width of the incident
phase space element $\delta k$ via
\begin{equation}
\label{eq:delk}
\delta k = \eta k_0.
\end{equation}
The volume of the incident phase space element is given by
\begin{equation}
\label{eq:vpsa}
V_{ps}=2\Delta k \delta k.
\end{equation}
Since the phase space volume for the incident and scattered neutrons
must be equal by Liouville's theorem, we have
\begin{eqnarray}
\label{eq:vpsb}
V_{ps}&=&2\Delta k \delta k \nonumber \\ &=&2\Delta k' \delta k'.
\end{eqnarray}

%
%
\begin{figure}[b]
\includegraphics[width=2.75in]{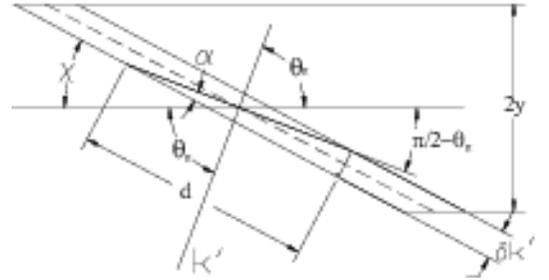}
\vspace*{3mm}
\caption{Magnified view of the Bragg-reflected phase space element.}
\label{pstmagnified}
\end{figure}

>From the geometry of the scattered phase space element we have
\begin{equation}
\label{eq:hb}
y=\Delta k' \sin\chi,
\end{equation}
which can be solved for $\Delta k'$ since we know $\chi$ and $y$ from
Eqs.~(\ref{eq:tanchi}) and (\ref{eq:h}) respectively,
\begin{equation}
\label{eq:deltakp}
\Delta k'={y\over{\sin\chi}}.
\end{equation}
Using Eqs.~(\ref{eq:vpsb}) and (\ref{eq:deltakp}) we can express
$\delta k'$ in terms of the known quantities,
\begin{equation}
\label{eq:delkp}
{\delta k'}={{V_{ps}\sin\chi}\over{2y}}.
\end{equation}
A magnified view of the scattered phase space element is illustrated
in Fig.~\ref{pstmagnified}.  In particular we show how the
monochromator (placed after the PST chopper) will intersect the phase
space element.  Note that at the optimal velocity this line will be
parallel to the long axis of the phase space element and the neutron
intensity gain will be a maximum.  From the geometry shown in
Fig.~\ref{pstmagnified}, the angle $\alpha$ is given by
$\alpha={\chi+\theta_g+{\pi\over 2}}$.  The projection of the
monochromator intersection with the phase space element is given by
$\min(d,2\Delta k')$ where $d$ is given by the geometry shown,
\begin{equation}
\label{eq:d}
d={{\delta k'}\over{\sin\alpha}}.
\end{equation}
The gain of the PST chopper is then given by the ratio of the final
phase space width (as seen by the monochromator) to the initial phase
space width defined in Eq.~(\ref{eq:delk}), and is expressed in terms
of the known quantities
\begin{equation}
\label{eq:gain} gain={{\min(d,2\Delta k')}\over{\eta k}}.
\end{equation}

\begin{figure}[t]
\includegraphics[width=2.75in]{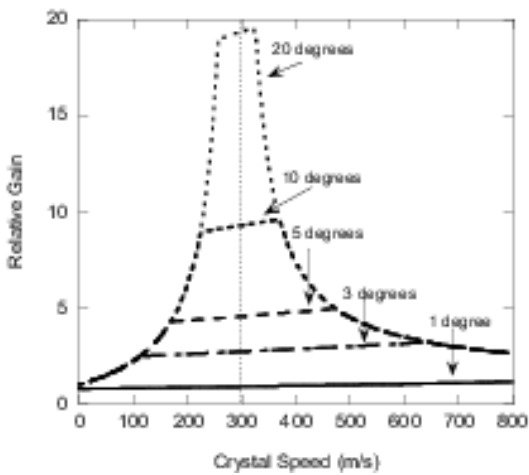}
\caption{ Relative gain in neutron flux as a function of crystal speed
  for several different values of mosaic spread based on
  Eq.~(\ref{eq:gain}).  The simulations assume a wavelength of
  $\lambda_0$.}
\label{pstangain}
\end{figure}

Fig.~\ref{pstangain} shows the calculated gain as a function of
crystal speed over a large range of mosaic spreads, $2\varepsilon
= 1^{\circ}, 3^{\circ}, 5^{\circ}, 10^{\circ}$ and $20^{\circ}$.
For mosaics of $3^{\circ}$ and larger, the gain increases rapidly
from one at low speeds, but then abruptly enters a linear region
at a mosaic-dependent speed.  At higher speed, the linear region
abruptly ends and the gain decreases smoothly from this point on.
For each mosaic value except $2\varepsilon=1^{\circ}$, the gain
curves lie on top of each other at sufficiently low and high
crystal speeds.  The only difference is the location of the linear
region that clips the top of the curve. The gains for this linear
region increase with increasing $2\varepsilon$, whereas the width
of the region shrinks. The linear region occurs when ${\rm
min}(d,2\Delta k')=2\Delta k'$ (see Fig.~\ref{pstmagnified}),
which happens to be satisfied for the case
$2\varepsilon=1^{\circ}$ over the entire crystal speed range
shown.  This happens because the mosaic of $1^{\circ}$ is too
small compared to the divergence of the incident beam.  The
maximum gain found here (19.5) occurs for the largest mosaic of
$20^{\circ}$ at a crystal speed of 320\,m/s.  Note however that
the reflectivity of the moving crystals has been assumed constant
for all cases of mosaic spread, an assumption that is incorrect
for real crystals.  In fact, at a constant thickness the
reflectivity of real crystals decreases with increasing mosaic
spread.  In addition, these calculations assume a $k$-independent,
or constant energy spectrum incident on the PST chopper, when in
fact the the cold source produces a Maxwell-Boltzmann neutron
energy distribution, which is further modified by the guide
coatings and filters downstream.  These and other simplifications
will cause the actual gain value to be less than that found in
this analytic calculation.

%
%

This analytical analysis quickly gives one a very good intuitive
feel for the relative gains one can expect from various
experimental configurations.  However, as mentioned above, the
calculations make several approximations, all of which tend to
overstate the gain, and which result in the chopped-off appearance
of the different gain curves.  The most important of these
approximations treats the scattered phase space element as a
parallelogram when in fact it is a rather complicated
crescent-shaped surface (see Fig.~\ref{pstphase}). Moreover, the
locus of final wavevectors accepted by the monochromator
collectively describe the surface of a sphere.  Thus, instead of
the gain being given by the intersection of a line with a
parallelogram, it should be given by the intersection of these two
more complicated curved surfaces.  (The word ``surface'' is
slightly misleading because both objects actually have non-zero
width along $k$.)  This will tend to round the shape of gain
curves and reduce the gain, particularly at high speeds, due to
the mismatch between the shape of the phase space element accepted
by the monochromator and that diffracted by the PST chopper.  To
account for these complex geometrical effects, as well as the
reflectivity dependence on mosaic spread and the true energy
spectrum seen by the PST chopper, Monte Carlo simulations of the
phase space transformation chopper have been performed.

The PST chopper is located immediately after the converging guide
described in Sec.~\ref{CG}, the supermirror coatings of which have a
critical angle of about twice that of $^{58}$Ni.  Based on
simulations, the divergence after this element was taken to be
2.8$^{\circ}$ in the horizontal plane and 4.4$^{\circ}$ in the
vertical direction.  The HOPG crystals were chosen to have a thickness
of 5\,mm.  The incident neutron energy spectrum was chosen to be a
65\,K Maxwell-Boltzmann distribution, in accord with measurements of
the flux from the NIST cold source, but truncated at 4\,\AA\ to
simulate the effect of a beryllium filter in the incident beam.  The
spectrum was also truncated at 10\,\AA\ because wavelengths longer
than this have no probability of being diffracted by the moving
crystal.  This distribution was then multiplied by the square of the
incident wavelength in order to account for the fact that the critical
angle for total external reflection is proportional to the wavelength.
The simulation included both horizontal and vertical mosaics, and the
velocity of the graphite crystal.  The program randomly chose the
incident wavelength, the angles of incidence (within limits given by
the divergences stated above), and the vertical orientation of the
crystallite involved in the Bragg reflection.  These parameters were
sufficient to determine the horizontal orientation required for
diffraction to occur.  In addition the reflectivity of the graphite
was included via the Bacon-Lowde equation for diffraction from ideally
imperfect crystals.~\cite{Bacon} The HOPG crystals used in the PST
chopper are most likely not all ``ideally imperfect.''  Thus even this
more realistic assumption still overestimates the reflectivity of the
HOPG crystals, and produces simulated gains larger than what are
actually observed.  All of these simulations have been performed using
silicon (111) crystals as the monochromator.

%
%
\begin{figure}[t]
\includegraphics[width=2.75in]{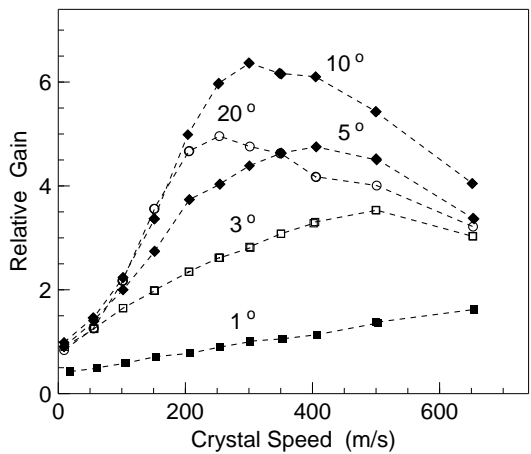}
\caption{ Simulations of the relative gain in neutron flux as a
  function of crystal speed for several different values of mosaic
  spread (numbers are FWHM).  The simulations assume a 5\,mm thick
  crystal and a wavelength of $\lambda_0$.}
\label{pstsimgain}
\end{figure}

Two-dimensional projections of simulated Bragg reflections from an
HOPG crystal having an isotropic $10^{\circ}$ mosaic are shown in
Fig.~\ref{pstphase} for three different crystal speeds.  A coordinate
system is used in which the $(x,y)$ plane coincides with the
horizontal scattering plane, and the $y$-axis is oriented antiparallel
to the (002) scattering vector.  The incident and final $k_x$ and
$k_y$ values of the diffracted neutrons are represented by individual
dots, while the value of $k_z$ is neglected.  This results in dots
whose total $k$ vectors seem abnormally short.  The reference values
of $k_x$ and $k_y$ are indicated by the solid lines.  Two effects are
evident in these simulations.  The first is that the phase space
element volume increases as the crystal velocity (oriented opposite
$k_x$) increases, in agreement with the analytic calculations.  This
is because the Bragg reflection takes place at a lower angle in the
Doppler frame.  The second is that the diffracted element tilts in
phase space as the crystal velocity is changed.  This tilt can be
optimized to maximize the number of neutrons that have the correct
energy to be backscattered from a Si \{111\} crystal.

The simulated flux gain, calculated using the same parameters as the
analytical gain shown in Fig.~\ref{pstangain}, is shown in
Fig.~\ref{pstsimgain}. As in the analytical calculation, the peak
intensity (relative to that obtained for a crystal velocity of zero)
is shown as a function of speed for mosaics $2\varepsilon =
1^{\circ}$, 3$^{\circ}$, 5$^{\circ}$, 10$^{\circ}$, and 20$^{\circ}$.
There are a number of similarities between the simulation and the
analytical calculation.  For mosaics of 3$^{\circ}$ or larger, the
relative intensity increases from about 1 to a broad maximum, before
decreasing again.  For the parameters chosen here, the maximum gain
($\sim 6$) occurs for a crystal with a 10$^{\circ}$ mosaic moving at
about 300\,m/s.  The results for a 20$^{\circ}$ crystal show a smaller
gain due to a lower reflectivity.  As in the analytical calculation,
the gain for the $1^{\circ}$ mosaic case increases linearly with
crystal speed, and has a relative value of less than one for zero
speed.

Phase-space transformation is an ideally suited method for boosting
the count rates on backscattering instruments because it increases the
divergence of the diffracted beam at the same time, thus alleviating
the mismatch in angular acceptance between the monochromator and
analyzer systems.  However, there are several considerations that
limit how far one can take the PST process.  For example, a PST
chopper using 10$^{\circ}$ mosaic crystals, which in our simulations
resulted in the highest gain, would produce a beam with a horizontal
divergence that is slightly larger than the 20$^{\circ}$ expected.  To
accept and use all of this divergence would require a monochromator
approximately 75\,cm long.  This would imply a heavy monochromator
assembly, and present serious technical challenges related to the
Doppler drive system.  Thus we compromised on a more reasonable
monochromator length of 52\,cm and a mosaic of about 7.5$^{\circ}$.
We also reduced the operational velocity of the PST chopper to
250\,m/s to make the forces on the composite disk more manageable and
increase the reliability.  Given these compromises, our simulations
predict that the gain from the PST should be about a factor of 5.
This can be compared to the gain of 6.7 calculated in the analytical
formulation using the same parameters, and agrees quite well with the
experimentally measured value of 4.2.

\section{Flight Time Offset Corrections}
\label{ftosc}

In this appendix we discuss the flight-time offset corrections
(FTOSC).  In particular we show that when the amplitude and velocity
of the monochromator are appreciable, parts of the spectrum undergo an
apparent shift.  It turns out that the elastic lineshape is affected
the most whereas scattering features at the largest energy transfers
are affected only marginally.

%
%
\begin{figure}[b]
\includegraphics[width=3.15in]{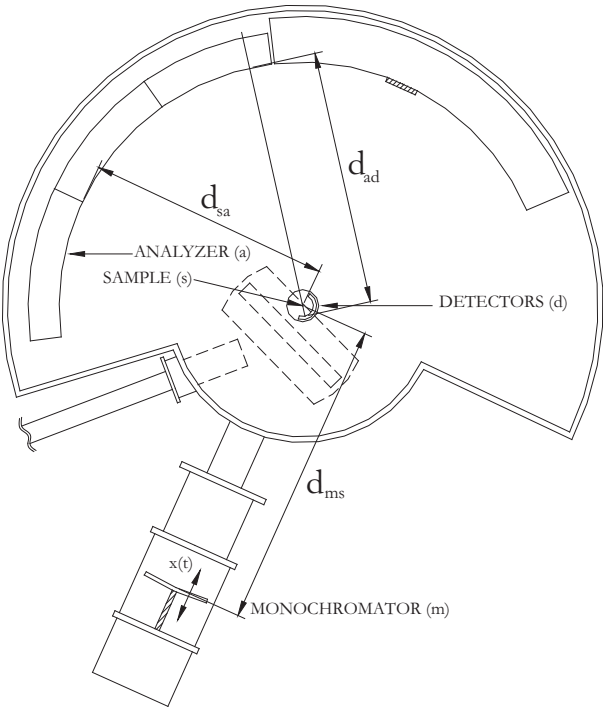}
\caption{ Schematic diagram of the relevant instrument geometry which
  provides the definition of terms used in the FTOSC.  }
\label{ftoscschem}
\end{figure}

A schematic of the relevant instrument geometry is presented in
Fig.~\ref{ftoscschem}.  The velocity of the neutrons reflected from
the Doppler monochromator is a time-dependent quantity, $v_n(t)$.  The
velocity of the monochromator, $v(t)$, modulates the reflected
velocities about the Bragg velocity, $v_0$, so that the neutrons that
strike the sample have a velocity $v_n(t)=v_0+v(t)$.  For a sinusoidal
monochromator motion with an amplitude of 4.5\,cm and a frequency of
25\,Hz, the maximum monochromator speed is 7.07\,m/s.  Thus the
neutron velocity varies between 632.7\,m/s and 637.9\,m/s.

The arrival time in the detectors, $t_d$, of the neutrons that
scattered from the monochromator at time $t$ is given by
\begin{equation}
  t_d=t+{d_{sad}\over v_0}+{d_{ms}+x(t)\over v_0+v(t)},
\end{equation}
where the sample-analyzer-detector distance is
$d_{sad}=d_{sa}+d_{ad}$, $d_{ms}$ is the sample-monochromator
distance, and $x(t)$ is the monochromator displacement (see
Fig.~\ref{ftoscschem}).  If we were to neglect the motion of the
monochromator in calculating the arrival time ($x(t)=0, v(t)=0$) we
obtain the zero-monochromator motion arrival time,
\begin{equation}
  t_0=t+{d_{sad}+d_{ms}\over v_0}.
\end{equation}
The actual arrival time, $t_d$, can be rewritten in terms of $t_0$ as follows,
\begin{equation}
  t_d=t_0+\delta t,
\end{equation}
where
\begin{equation}
  \delta t={d_{ms}\over v_0}{{x\over d_{ms}}-{v\over
      v_0}\over{1+{v\over v_0}}}.
\end{equation}
This term, $\delta t$, is the flight-time offset correction term.
Note that it only becomes important for large values of the quantities
$x/d_{ms}$ and $v/v_0$.  The large-amplitude/high-frequency operation
of the HFBS Doppler monochromator yields non-negligible values for
$\delta t$.  To appreciate the magnitude of $\delta t$ under routine
operating conditions on HFBS, let us assume that the monochromator
frequency is 17.7\,Hz (corresponding to a dynamic range of $\pm 33 \mu
eV$).  At maximum monochromator velocity, $v=5$ m/s and $x=0$.  At
zero velocity, $x=0.045$\,m.  Using $d_{ms}=2.25$\,m,
$v_0=630.8$\,m/s, and $d_{sad}=4.12$\,m, we obtain $\delta t =
-28$\,$\mu$s at maximum monochromator velocity, while $\delta
t=71$\,$\mu$s at zero monochromator velocity.  Since the cam that
moves the monochromator rotates at a uniform rate, constant cam
intervals are also constant time intervals.  With constant time bins
of 20\,$\mu$s (the smallest time bin in the HFBS data acquisition
configuration), parts of the spectrum can be shifted by as much as
three channels.  However, since $\delta t$ varies, different portions
of the spectrum will be affected differently.

One is usually interested in a spectrum in energy, not
time-of-arrival.  Therefore we can also calculate the implications for
the energy when the FTOSC time is used in the data acquisition and
when it is ignored.  When the motion of the monochromator is ignored
in the time-of-arrival in the detectors, the time-dependence of the
energy of the neutron after reflection from the monochromator is given
simply by
\begin{equation}
  E_n^{no ftosc}=E(t_0).
\end{equation}
If we take into account the FTOSC term then
\begin{equation}
  E_n^{ftosc}=E(t_0+\delta t).
\end{equation}
To first order we may expand this expression out in $\delta t$,

\begin{eqnarray}
  E_n^{ftosc} & = & E(t_0)+\delta t\frac{\partial E}{\partial
    t}\left. \right|_{t_0} \\ & = & E+\delta E.
\end{eqnarray}
Thus for small values of $\delta t$, the spectrum will be shifted by
an amount proportional to $\delta t$.  In our case with $f_{Dop}=17.7$
Hz, we find $\delta E=0.3 \mu $eV, or roughly one third of the
instrumental resolution.  When $E$ is a maximum,
$\partial{E}/\partial{t}$ is zero, resulting in no shift.

The spectral lineshape will also be distorted since the Jacobian in
the nonlinear transformation is not unity, although this is a much
weaker effect than the shift.  As is the case with the shift, the
distortion is most pronounced at the elastic peak position, where the
monochromator velocity is zero, and smallest when the monochromator
speed is maximum.


\begin{references}




\bibitem{Maier} H. Maier-Leibnitz, Nukleonik {\bf 8}, 61 (1966).

\bibitem{Alefeld} B. Alefeld, Bayer. Akademie der Wissenschaften,
  Math. Naturwiss. Klasse {\bf 11}, 109 (1966).

\bibitem{Bottom} V. E. Bottom, Anais da Academia Brasileira
  Ci\^eneias {\bf 37}, 407 (1965).

\bibitem{darwin} C. G. Darwin, Phil. Mag. {\bf 27}, 315 and 675
  (1914); P. P. Ewald, Z. Phys. {\bf 30}, 1 (1924).

\bibitem{Birr} M. Birr, A. Heidemann, and B. Alefeld, Nucl.  Instr.
  Methods {\bf 95}, 435 (1971).

\bibitem{si220} G. Basile, A. Bergamin, G. Cavagnero, E. Vittone, and
  G. Zosi, Phys. Rev. Lett. {\bf 72}, 3133 (1994).

\bibitem{Cook} J. C. Cook, W. Petry, A. Heidemann, and J-F.
  Barth\'{e}lemy, Nucl. Instr. and Meth. {\bf A312}, 553 (1992).

\bibitem{illyb} Yellow Book ILL.

\bibitem{GeNe98} P. M. Gehring and D. A. Neumann, Physica B {\bf
    241--243}, 64 (1998).

\bibitem{Williams} R. E. Williams, J. M. Rowe, and P. Kopetka,
  Proceedings of the International Workshop on Cold Moderators for
  Pulsed Neutron Sources, Argonne National Laboratory, Sept. 29 - Oct.
  2, 1997, J. M. Carpenter and E. B. Iverson, editors, pp. 79-86.



\bibitem{Chuck} C. F. Majkrzak and J. F. Ankner, in {\it Neutron
    Optical Devices and Applications}, C. F. Majkrzak and J. L.  Wood,
  eds.  SPIE Proc. Vol. 1738, (SPIE, Bellingham, WA, 1992) p.  150.

\bibitem{deGraaf} L. A. de Graaf, IRI report 132-82-02/1, (1984).

\bibitem{Glinka} C. J. Glinka, private communication.

\bibitem{update} This number should be multiplied by 1.8 to account
  for the increased cold neutron flux provided by the new hydrogen
  cold source described in Sec.~\ref{prospects}.

\bibitem{Anderson} I. S. Anderson, in {\it Thin-Film Optical Devices:
    Mirrors, Supermirrors, Multilayer Monochromators, Polarizers and
    Beam Guides}, ed. C. F. Majkrzak, Proc. SPIE {\bf 983}, 84 (1989).

\bibitem{Copley} J. R. D. Copley, J. Neutron Research {\bf 1}, 21
  (1993).

\bibitem{Jeremy} J. C. Cook, private communication.

\bibitem{Schelten} J. Schelten and B. Alefeld, in Proc. Workshop on
  Neutron Scattering Instrumention for SNQ, edited by R. Scherm and H.
  H.  Stiller, report J\"{u}l-1954 (1984); G. S. Bauer and R. Scherm,
  Physica B {\bf 136}, 80 (1986).

\bibitem{Shapiro} S. M. Shapiro and N. J. Chesser, Nucl. Instr. and
  Meth. {\bf 101}, 183 (1972).

\bibitem{Shull} C. G. Shull and N. S. Gingrich, J. Appl. Phys.  {\bf
    35}, 678 (1964); B. Buras and T. Giebultowicz, Acta. Cryst.  {\bf
    A 28}, 151 (1972); D. Bally, E. Tarina, and N. Popa, Nucl. Instr.
  and Meth. {\bf 127}, 547 (1975).

\bibitem{Stoica} A. D. Stoica and M. Popovici, J. Appl. Cryst.  {\bf
    22}, 448 (1989).

\bibitem{etch} The etching was done in a 10\%\ solution of HF acid.

\bibitem{Bacon} G. E. Bacon and R. D. Lowde, Acta Cryst. {\bf 1},
  303 (1948).

\bibitem{Mihai} M. Popovici, private communication.

\bibitem{Meyer2000} A. Meyer et al., to be published.

\bibitem{Meyer2002} A. Meyer, H. Schober, D.\,B. Dingwell, Europhys. Lett. (in press).

\bibitem{Fillaux} F. Fillaux, C. J. Carlile, G. J. Kearley, M. Prager,
  Physica B {\bf 202}, 302 (1994).

\bibitem{Mukhopadhyay} R. Mukhopadhyay, C. J. Carlile, R. N. Silver,
  Physica B {\bf 174}, 546 (1991).

\bibitem{Sivia2} D. S. Sivia and C. J. Carlile, J. Chem. Phys. {\bf
    96}, 170 (1992).

\bibitem{Sivia1} D.S. Sivia, Physica B {\bf 202}, 332 (1994).

\bibitem{Springer77} T. Springer, in {\it Dynamics of Solids and
    Liquids by Neutron Scattering}, eds. S.W. Lovesey and T. Springer,
  (Springer-Verlag, Berlin Heidelberg, 1977) p. 284.

\bibitem{newsource} R. E. Williams and J. M. Rowe, Physica B {\bf
    311}, 117 (2002); For current performance information refer to
  http://www.ncnr.nist/gov/coldgains/.

\bibitem{Boni} P. B\"{o}ni, private communication.


\end{references}
\end{document}